\documentclass[acmtog, screen, nonacm]{acmart}

\renewcommand{\footnotetextcopyrightpermission}[1]{}
\setcopyright{none}
\acmBooktitle{}
\acmConference{}
\acmISBN{}
\acmDOI{}

\makeatletter
\def\@folioblob{}
\def\@runningfoot{}
\makeatother

\usepackage{booktabs}
\usepackage{bbm}
\usepackage{nicefrac}
\usepackage{algorithm}
\usepackage{algpseudocode}
\usepackage{algorithmicx}
\usepackage{enumitem}
\usepackage[normalem]{ulem}
\usepackage{empheq}
\usepackage{cleveref}
\usepackage{booktabs}
\usepackage{tabularx}
\usepackage{array}
\usepackage{amsmath}
\definecolor{codegreen}{rgb}{0,0.6,0}
\definecolor{codegray}{rgb}{0.5,0.5,0.5}
\definecolor{codepurple}{rgb}{0.58,0,0.82}
\definecolor{backcolour}{rgb}{0.96,0.96,0.98}
\usepackage{xcolor}

\theoremstyle{definition}

\usepackage{listings}
\usepackage[normalem]{ulem}

\setcopyright{none}
\setcitestyle{square}

\title{MPM Lite: Linear Kernels and Integration without Particles} 

\author{Xiang Feng}\authornote{Equal contribution.}
\affiliation{
\institution{University of California, Los Angeles \& University of California, San Diego}
\country{USA}
}
\email{xfeng.cg@gmail.com}

\author{Yunuo Chen}
\authornotemark[1]
\affiliation{
\institution{University of California, Los Angeles}
\country{USA}
}
\email{yunuoch@gmail.com}

\author{Chang Yu}
\authornotemark[1]
\affiliation{
\institution{University of California, Los Angeles}
\country{USA}
}
\email{g1n0st@live.com}

\author{Hao Su}
\affiliation{
\institution{University of California, San Diego}
\country{USA}
}
\email{academic@haosu.ai}

\author{Demetri Terzopoulos}
\affiliation{
\institution{University of California, Los Angeles}
\country{USA}
}
\email{dt@cs.ucla.edu}

\author{Yin Yang}
\affiliation{
\institution{University of Utah}
\country{USA}
}
\email{yangzzzy@gmail.com}

\author{Joe Masterjohn}
\affiliation{
\institution{Toyota Research Institute}
\country{USA}
}
\email{joe.masterjohn@tri.global}

\author{Alejandro Castro}
\affiliation{
\institution{Toyota Research Institute}
\country{USA}
}
\email{alejandro.castro@tri.global}

\author{Chenfanfu Jiang}
\affiliation{
\institution{University of California, Los Angeles}
\country{USA}
}
\email{chenfanfu.jiang@gmail.com}

\usepackage{amsmath}

\usepackage{multirow}
\usepackage{caption}
\usepackage{subcaption,bm}
\usepackage{ stmaryrd }
\algdef{SE}[DOWHILE]{Do}{doWhile}{\algorithmicdo}[1]{\algorithmicwhile\ #1}
\AtBeginDocument{%
  \providecommand\BibTeX{{%
    \normalfont B\kern-0.5em{\scshape i\kern-0.25em b}\kern-0.8em\TeX}}}

\begin{document}

\begin{abstract}
In this paper, we introduce MPM Lite, a new hybrid Lagrangian/Eulerian method that eliminates the need for particle-based quadrature at solve time. Standard MPM practices suffer from a performance bottleneck where expensive implicit solves are proportional to particle-per-cell (PPC) counts due to the the choices of particle-based quadrature and wide-stencil kernels. In contrast, MPM Lite treats particles primarily as carriers of kinematic state and material history. By conceptualizing the background Cartesian grid as a voxel hexahedral mesh, we resample particle states onto fixed-location quadrature points using efficient, compact linear kernels. This architectural shift allows force assembly and the entire time-integration process to proceed without accessing particles, making the solver complexity no longer relate to particles. At the core of our method is a novel stress transfer and stretch reconstruction strategy. To avoid non-physical averaging of deformation gradients, we resample the extensive Kirchhoff stress and derive a rotation-free deformation reference solution, which naturally supports an optimization-based incremental potential formulation. Consequently, MPM Lite can be implemented as modular resampling units coupled with an FEM-style integration module, enabling the direct use of off-the-shelf nonlinear solvers, preconditioners, and unambiguous boundary conditions. We demonstrate through extensive experiments that MPM Lite preserves the robustness and versatility of traditional MPM across diverse materials while delivering significant speedups in implicit settings and improving explicit settings at the same time. Check our project page at \url{https://mpmlite.github.io}.
\end{abstract}



\begin{teaserfigure}
    \includegraphics[width=\textwidth]{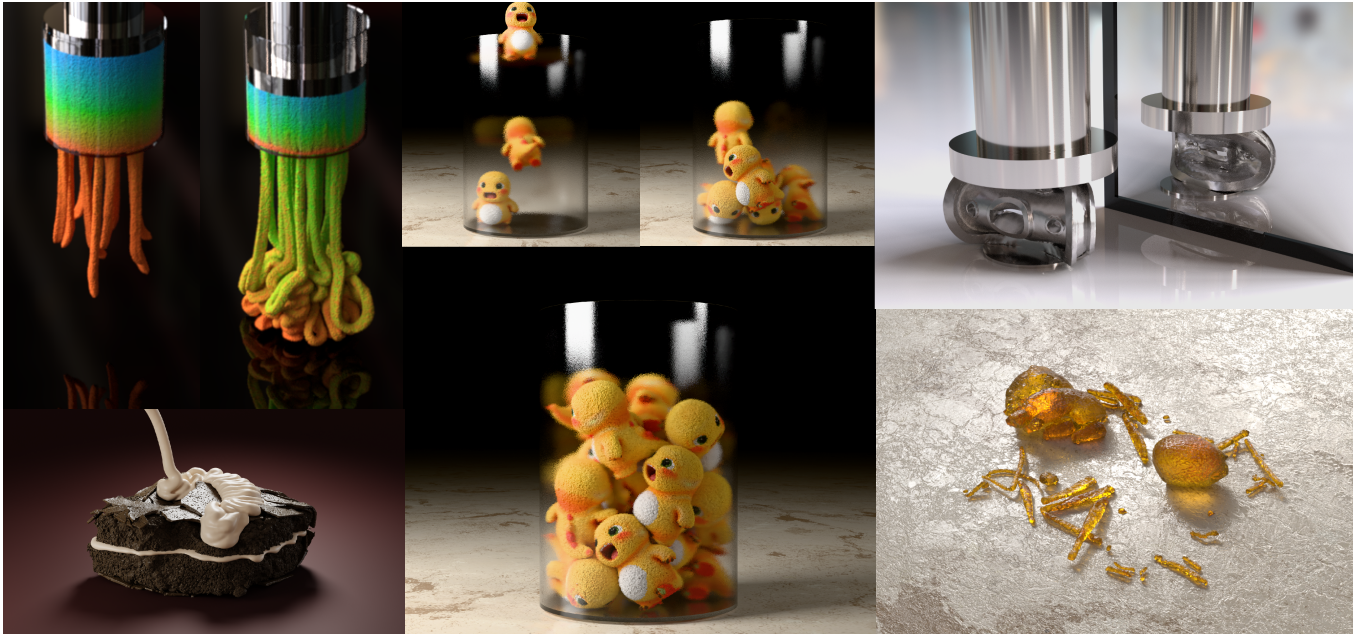}
  \caption{\textbf{Versatile material simulation using MPM Lite.} From left to right and top to bottom: noodles extruded through a cylindrical mold; cream spread over the surface of a brownie; soft stuffed toys dropped into a glass container; a stiff aluminum wheel deformed by a hydraulic press; and an ant-shaped candy impacting the ground and fracturing. Each example demonstrates the versatility, robustness, and scalability of the MPM Lite simulator across a wide range of material behaviors.}
  \label{fig:teaser}
\end{teaserfigure}

\maketitle

\section{Introduction}

The Material Point Method (MPM) has become a workhorse for graphics and computational mechanics, enabling robust simulation of large deformation, fracture, and multi-phase flow. Despite its popularity, widely used variants inherit two structural burdens. 
First, modern MPM practice typically relies on higher-order B-spline transfer kernels to mitigate cell-crossing instabilities that arise with linear bases. While effective, these kernels complicate massive parallelization, make implicit solvers expensive (due to wide stencils), and render boundary-condition enforcement ambiguous because B-splines lack the Kronecker-delta property at grid nodes. 
Second, forces are integrated through particle quadrature. This yields a clean derivation from the weak form, but the computational cost scales directly the particle-per-cell (PPC) count. In implicit settings, every gradient or Hessian vector product evaluation triggers a full grid-to-particle-to-grid (G2P2G) loop over all particles, creating a performance cliff as resolution or PPC increases.

It is common to describe MPM as FEM in disguise: particles as quadrature samples, serving as elements; the background grid as reference space, serving as vertex mesh. Yet standard FEM has none of the above liabilities. FEM employs fixed (and typically Gaussian) quadrature at geometry-aware locations and compact, nodal shape functions with clean boundary semantics. The discrepancy between the MPM paradigm and the FEM practice motivates a simple question: Must MPM particles be quadrature?

We argue no. In the closest MPM-related scheme, FLIP/APIC fluids, particles act primarily as markers that carry the the kinematic state between grid solves and serve as topology trackers that help advection. Solids, of course, require strain, stress, and material histories; their particles need to remember more. But using them as integration points at solver time is merely one choice of approximating stress integrals arising from the variational weak form. If we can let particles track, sample, and faithfully resample state onto a small, grid-aligned quadrature, we can move force assembly and implicit integration fully to the grid, avoiding repeated, PPC-proportional particle loops. This perspective also resolves the B-spline issue. Cell crossing was historically a particle-grid transfer pathology. If quadrature does not reside on particles, the force assembly becomes less sensitive to high-frequency particle motion. With an appropriate resampling scheme, we can use compact, linear kernels for communication while maintaining stability by construction.

Based on the above arguments we introduce MPM Lite: a new MPM-like hybrid Lagrangian/Eulerian discretization scheme that downplays particle quadrature at solver time. Conceptually, we treat the Cartesian grid as a voxel hexahedral mesh; we resample particle fields to fixed-location quadrature points on the mesh with compact and efficient linear kernels, and we perform force assembly and the entire implicit (or explicit) time integration without accessing particles, returning to particles only for advection and constitutive model updates. The main components that form the core of MPM Lite are
\begin{enumerate}
\item A linear-kernel transfer scheme that communicates momentum and stress between particles and the grid. Our velocity communication is second-order consistent to B-Spline APIC and our stress communication avoids cell-crossing instability by construction.
\item A spatial-temporal force integration viewpoint that does not require the participation of MPM particles. The computational focus is concentrated to elements rather than particles.
\item A custom optimization-based incremental potential formulation that utilizes a rotation-free stretch reference in the updated-Lagrangian linearization for isotropic materials. Our scheme is compatible with common material models and existing MPM variants.
\end{enumerate}
By removing particles from the integration (and only use them for information tracking), MPM Lite resembles one-point hexahedral finite elements on the voxel mesh: solver complexity no longer relates to particles, existing FEM nonlinear solvers, preconditioners, and subspace integration methods apply out of the box, and boundary conditions become unambiguous. MPM Lite can be easily implemented as independent particle resampling modules and an expandable FEM integration module. It achieves $1.88\times$ speedup over explicit MPM, and $15.9\times$ speedup over implicit MPM, on practical elastoplastic materials.

\section{Related Work}
\label{sec:related}

MPM was introduced by \citet{Sulsky1995} as a particle-grid hybrid for large‑deformation solids and has become popular in computer graphics and mechanics; see \citet{jiang2016material} for a comprehensive tutorial perspective. Two structural issues motivate much of the literature: (i) \emph{cell‑crossing artifacts} that arise when material points traverse element boundaries under $C^0$ multilinear bases and (ii) the \emph{cost and noise} of particle‑centric quadrature during force assembly \cite{Steffen2008}. One influential direction of research enlarges or convects particle domains to smooth particle-grid transfers. GIMP performs a convolution of grid shape functions with a finite particle characteristic function to suppress transfer discontinuities \cite{Bardenhagen2004}. CPDI generalizes this idea by allowing the particle domain to deform with the flow (e.g., parallelograms in 2D), improving accuracy under large shear and rotation \cite{Sadeghirad2011}; CPDI2 enriches corner sampling and better handles weak discontinuities at interfaces \cite{Sadeghirad2013}. Rather than widening supports, DDMP corrects shape‑function gradients to mitigate discontinuities directly \cite{Zhang2011}. A more recent kernel design, CK‑MPM, introduces compact dual‑grid stencils that lower per‑particle scatter cost while retaining particle quadrature \cite{liu2025ckmpm}. A largely orthogonal research direction replaces piecewise‑linear tent functions with smoother grid bases so that forces vary continuously as particles cross cells: B‑spline MPM adds $C^1$-$C^2$ continuity and reduces quadrature error \cite{Steffen2008}, with follow‑up work combining spline interpolation and tailored quadrature to improve robustness \cite{Gan2018}. Isogeometric MPM (IGA‑MPM) adopts NURBS‑style bases to attain higher‑order convergence and exact geometry representation \cite{Moutsanidis2020}. While these strategies markedly reduce artifacts, they also broaden stencils and complicate boundary conditions and implicit assembly--trade-offs that motivate our decision to keep compact $Q1$ kernels while addressing integration elsewhere.

A second and complementary series of work relocate integration from moving particles to fixed grid points, blurring the boundary with updated‑Lagrangian FEM. In geomechanics, \citet{AlKafaji2013} proposed mixed schemes that use standard Gauss quadrature in fully filled cells and revert to particle quadrature near interfaces. Improved MPM formulations assemble the weak form entirely on the background grid while reconstructing kinematics via MLS transfers \cite{Sulsky2016}; high‑order variants pair fixed Gauss rules with smooth bases to further reduce integration noise \cite{Tielen2017, Gan2018}. These approaches inherit FEM's clean sparsity and boundary semantics but hinge on accurate, stable resampling between particles and quadrature points. We follow this grid‑quadrature lineage yet make two deliberate choices for graphics practice: we resample to cell‑center quadrature using compact $Q1$ kernels and perform force/Jacobian assembly entirely on the grid, so solver cost becomes independent of particles‑per‑cell while particles retain their role as history carriers. This design relates to and contrasts with staggered‑grid MPM \cite{liang2019efficient}, which reallocates FLIP/PIC transfers across interleaved grids for explicit MPM and evolves stress in rate form; our method targets APIC-style kernels, deformation‑gradient‑based hyperelastoplasticity, and fully implicit solves with finite element stencils.

Within graphics, MPM matured into a versatile engine for complex materials and topological change. The snow system established robust large‑deformation and phase‑change behavior in production settings \cite{stomakhin2013snow,stomakhin2014augmpm}. Transfer design proved central to visual fidelity: APIC augments PIC/FLIP with local affine velocity modes to preserve angular momentum and reduce dissipation \cite{jiang2015apic}, and subsequent variants refine particle-grid exchange via polynomial carriers, power‑weighted moments, and impulse‑centric formulations \cite{fu2017polypic,qu2022powerpic,sancho2024impulse}. Our communication layer aims to match APIC‑level fidelity while retaining \emph{linear} kernels; by moving integration off particles, we keep stencils compact and decouple solver complexity from PPC. The community simultaneously broadened constitutive modeling and coupling. Drucker-Prager elastoplasticity yields convincing granular flow \cite{klar2016sand}; multi‑species porous sand-water and particle‑laden flows capture debris and sediment dynamics \cite{tampubolon2017multispecies,gao2018particleladen}; and thin‑shell MPM supports frictional contact for sheet‑like structures \cite{guo2018shells}. Robust two‑way coupling advanced via MLS‑MPM with displacement discontinuities and rigid interactions \cite{hu2018mlsmpm} and via interface‑aware quadrature for non‑sticky fluid-solid coupling \cite{fang2020iqmpm}. Stability and performance have likewise seen sustained investment: implicit formulations for non‑equilibrated viscoelastic/elastoplastic solids enable large time steps \cite{fang2019sillyrubber}; GPU‑oriented kernels and hierarchical time integration accelerate stepping and linear solves \cite{gao2018gpumpm,wang2020hot}; and multi‑GPU designs scale to production problem sizes \cite{wang2020multigpu}. Differentiable pipelines enable inverse design and control in soft robotics and beyond \cite{hu2019chainqueen,hu2020difftaichi}, while damage/fracture models extend MPM beyond smooth elasticity \cite{wolper2019cdmpm,wolper2020anisompm}. Production reports emphasize robustness, predictable performance, and artist control \cite{klar2017production}. Our contribution is designed to dovetail with these priorities: fixed, FEM‑like sparsity and clean boundary semantics ease integration into existing nonlinear solvers and preconditioners; compact transfers preserve the accuracy expected by APIC‑class schemes; and to‑element assembly aligns naturally with differentiable and optimization‑driven workflows common in modern graphics pipelines.

\section{Unloading/Loading Particle Information}

We use subscript $p$ to denote particle quantities, $c$ to denote quadrature quantities which without loss of generalizability are chosen to be located at cell centers in this paper, $i$ to denote grid node quantities. This section discusses how to communicate discrete fields across these locations purely using linear kernels. Unlike traditional MPM where particles are still involved during the integration of forces, here particles act as transient couriers that unload to the grid for integration and reload afterward. We use the term ``load'' and ``unload'' to highlight this unique feature.

\subsection{Kinematic Transfers}
Given particles with mass $m_p$, rest volume $V_p$, position $x_p$, velocity $v_p$, deformation gradient $F_p$, and a matrix $G_p$ encoding velocity gradient information, we \emph{unload} kinematic quantities onto cell centers through a multilinear kernel $w_{cp} = N_c(x_p)$:
\begin{align}
m_c^n & = \sum_p w_{cp}\, m_p \label{eq:p2c-mV} \\
(mv)_c^{n} & = \sum_p w_{cp}\, m_p ( v_p^{n} \;+\; G_p^{n} (x_c - x_p^{n}) ), \label{eq:p2c-mv-affine}\\
G_c^{n} & = \sum_p w_{cp}\, m_p\, G_p^{n}/m_c^n, \label{eq:p2c-A-sum}
\end{align}
where mass and momentum transfers match APIC. The mass weighting in  velocity gradient extrapolation allows us to closely match the local velocity field reconstructed with quadratic B-splines up to an error of $O(\Delta x^2)$ (where $\Delta x$ is the grid spacing); see \cref{sec:kinematic_transfer_error}.

Unlike APIC/MLS-MPM, which reconstructs a per-particle affine velocity based on moment matrices and outer products of velocities with particle-grid offsets, our method keeps the velocity gradient directly on quadratures; \emph{loading} back to particles is just straightforward interpolation:
\begin{align}
v_p^{n+1} &= \sum_c w_{cp} v_c^{n+1}, \label{eqn:vc2p}\\
G_p^{n+1} & = \sum_c w_{cp}\ G_c^{n+1}. \label{eqn:Gc2p}
\end{align}
With $v_p^{n+1}$ we advect particle locations $x_p^{n+1}=x_p^{n}+\Delta t v_p^{n+1}$ and with $G_p^{n+1}$ we update particle deformation gradient as $F_p^{n+1} = (I + \Delta t G_p^{n+1}) F_p^{n}$.

\begin{figure}
    \includegraphics[width=\linewidth]{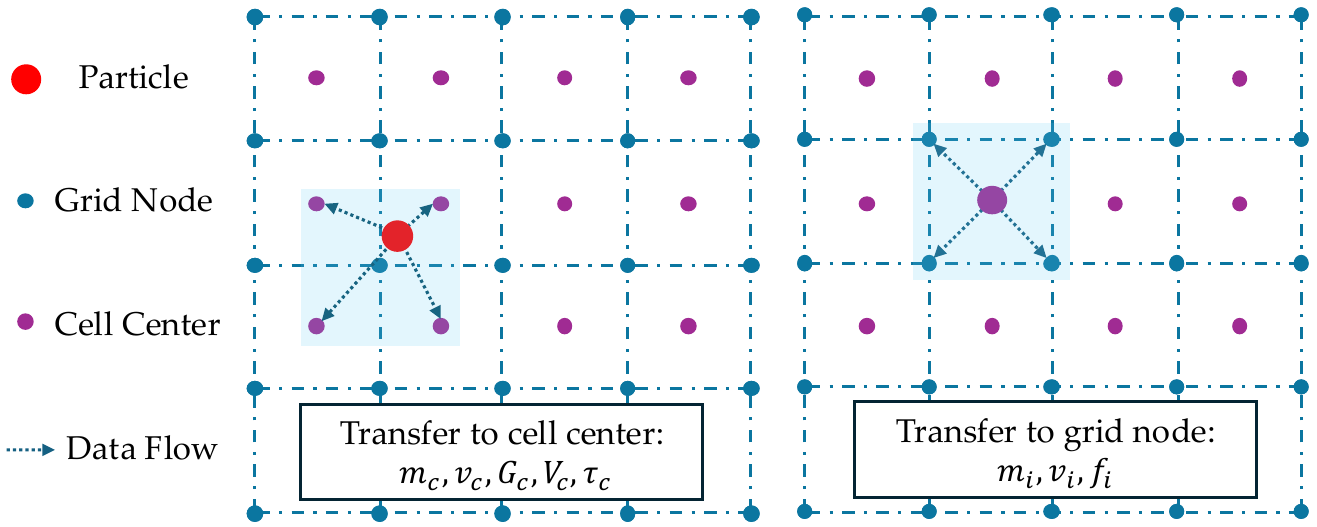}
  \caption{\textbf{Particle-to-center transfer and center-to-grid transfer}. In the particle-to-center transfer stage, particles unload their mass and momentum to the cell centers. Each cell center additionally accumulates volume and Kirchhoff stress. In the subsequent center-to-grid transfer, grid nodes gather mass, velocities, and forces from neighboring cell centers through a multilinear kernel. Owing to the constant kernel weights and purely gather-based formulation, this transfer can be parallelized without race conditions.}
  \label{fig:transfer_scheme}
\end{figure}

\subsection{Stress Transfers} \label{sec:stress-transfer}

To enable strain/stress computations on the grid, a tempting idea is to just unload the deformation gradient $F$ with mass or volume weighting. Averaging deformation gradients has indeed be done in previous work, often in the context of adaptively resampling particles \cite{yue2015continuum,gao2017agimp}. This, however, is not objective or physically meaningful: $F$ is neither intensive nor extensive, so mixing distinct particle rotations and stretches into a single cell value corrupts the stress state and stiffness. Indeed, we found that averaging $F$ leads to non-physical forces and solver instabilities.

The choice of strain/stress fields represented by quadrature points can be inspired by the reason quadrature points exist. In the case of MPM, quadratures are for approximating the force term in a Galerkin weak form \cite{jiang2016material}. In particular, the continuous elastic force under the action of an arbitrary test function $q(\cdot,t^n):\Omega^n\rightarrow\mathbb{R}^{d}$ ($d=2 \text{ or } 3$) is given by 
$-\int_{\Omega^n} \sigma(x,t^n) : \nabla q(x,t^n) d x$, where $\sigma = \tau/\text{det}(F)$ is Cauchy stress. One would pick proper shape functions as $q$, and evaluate the integral either exactly or with quadrature points. The Riemann sum  $\int_{\Omega^n} \sigma   =\int_{\Omega^0} \tau\approx \sum_c V_c \tau_c$ implies that the Kirchhoff stress content, $V\tau$, is clearly a proper extensive quantity whose net contributions should be preserved across any resampling. Accordingly, we perform
\begin{align}
V_c^n &= \sum_p w_{cp} V_p \\
\tau_c^n &= \sum_p w_{cp} (V_p \tau_p^{n})/V_c^n, \label{eqn:vtau}
\end{align}
which satisfies $\sum_c V_c^n\tau_c^n=\sum_p V_p \tau_p^n$. Here the symmetry of $\tau$ can be used to minimize practical computation and storage. Unlike kinematic variables, we neither load stress back to nor store it on particles since we already track $F_p$, and $\tau_p^{n}$ can be computed in place.

\subsection{Becoming Finite Elements}
\label{sec:becoming_fem}

After unloading particles, we have $\{m_c^n, v_c^n, G_c^n, V_c^n, \tau_c^n\}$. Despite the lack of deformation gradients, this forms complete ingredients for us to perform time integration on the lattice through an updated Lagrangian hexahedral FEM view. For simplicity, we adopt mass lumping and initiate data on grid node $i$ through a multilinear kernel $w_{ic} = N_i(x_c) \equiv 1/2^{d}$:
\begin{align}
m_i^{n} &= \sum_c w_{ic} m_c^n, \\
(mv)_i^{n} &=  \sum_c w_{ic} m_c^n ( v_c^{n} + G_c^{n} (x_i - x_c)),
\end{align}
followed by $v_i^{n}=(mv)_i^{n}/m_i^{n}$. The constant weight allows them to be implemented as a nodal gather operation with no race condition, as illustrated in Fig.~\ref{fig:transfer_scheme}.  The integrated $v_i^{n+1}$ (\autoref{sec:Incremental Potential}) along with its gradient are sampled back at the quadratures through:
\begin{align}
v_c^{n+1} &= \sum_i w_{ic}\, v_i^{n+1}, \\
G_c^{n+1} &= \sum_i v_i^{n+1} \otimes \nabla w_{ic}(x_c),
\end{align}
where  $\nabla w_{ic} = \nabla N_i(x_c) = (x_i-x_c)/(2^{d-2} \Delta x^2)$. $v_c^{n+1}, G_c^{n+1}$ are then loaded to the particles through \cref{eqn:vc2p,eqn:Gc2p}.

\subsection{Error Analysis} \label{sec:kinematic_transfer_error}

It is worth looking into the difference between our scheme and the more traditional quadratic B spline APIC scheme widely used in computer graphics. We prove in \autoref{appendix:v} that under reasonable Lipschitz smoothness assumptions on local velocity fields, the discrepancy in velocity and velocity gradient between the two schemes are minimal: they are both $O(\Delta x^2)$ with respect to the grid spacing $\Delta x$.

\section{Incremental Potential Formulation}\label{sec:Incremental Potential}

Explicit time integration for MPM Lite can be easily done through imposing internal forces
\begin{equation}
f_i^n = -\sum_c V_c^n \tau_c^n \nabla w_{ic} \label{eqn:explicitforce}
\end{equation}
on grid nodes and advancing their velocities with $v_i^{n+1} = v_i^n + \Delta t f_i^n/m_i^n$. In this section, we focuses on what to do for implicit integration.

\subsection{Problem with Implicit Stress}

Without loss of generality, we target an optimization-based time integration formulation for backward Euler. A natural temptation -- because we already carry cell-center stress -- is to keep stress in rate form, and make the Kirchhoff stress \(\tau\) an implicit function of the nodal velocity via an objective rate. To make this concrete on our center quadratures, let $D_c(v):=\operatorname{sym} G_c(v)$, $W_c(v):=\operatorname{skw} G_c(v)$, and for illustrational purpose  use a small‑strain elastic modulus \(\mathbb C\) evaluated at the beginning of the step. A backward-Euler Jaumann update that is implicit in \(v\) are explicit in \(\tau\) is
\begin{equation}
\label{eq:BE-Jaumann-semi}
\tau^{n+1}_c(v)
=\tau^n_c+\Delta t\Big(\,\mathbb C : D_c(v)\;+\;W_c(v)\,\tau^n_c-\tau^n_c\,W_c(v)\Big).
\end{equation}
Implicit nodal internal forces are then
\begin{equation}
\label{eq:fint-center}
f^{\mathrm{int}}_i(v) \;=\; -\sum_{c} V_{c}^n\,\tau^{n+1}_c(v)\,\nabla w_{ic}.
\end{equation}
The Fréchet derivative of \(\tau^{n+1}_c\) with respect to the nodal velocities follows by linearizing \eqref{eq:BE-Jaumann-semi}. Due to the contribution of the skew-adjoint term, the Jacobian \(\partial f^{\mathrm{int}}/\partial v\) is generically non‑symmetric despite a symmetric $\mathbb C$, and does not derive from any elastic potential energy, making it unsuitable for  optimization‑based implicit solvers. A fully implicit variant that also places \(\tau^{n+1}\) inside the Jaumann co‑rotation leads to the same problem. 

\subsection{A Rotation-Free Stretch Reference Solution}
\label{subsec:rotfree}

With this in mind we take the standard velocity‑primary incremental potential viewpoint and specialize it to our center quadrature. Let's for a second pretend we did manage to transfer deformation gradients  from particles to quadratures to give us an $F_c^n$ field (although as discussed in \autoref{sec:stress-transfer} this to-our-knowledge remains to be truly solved), a center trial deformation is then defined multiplicatively by updated Lagrangian:
\begin{equation}
F_c(v)=\bigl(I+\Delta t\,G_c(v)\bigr)\,F_c^{\text{base}}.
\label{eq:Ftrial-rotfree}
\end{equation}
Given a hyperelastic density $\psi(F)$ and quadrature rest volumes $V_c^n$, our time step solves the optimization problem \cite{GSSJT15} (with external forces omitted for simplicity)
\begin{equation}
\min_{\,v}\;
\underbrace{\sum_i^n \tfrac{1}{2}\,m_i\,\|v_i-v_i^n\|^2}_{\text{inertia}}
\;+\;
\underbrace{\sum_c V_{c}^n\,\psi\,\bigl(F_c(v)\bigr)}_{\text{discrete elastic energy}}.
\label{eq:incpot-rotfree}
\end{equation}
Solving \eqref{eq:incpot-rotfree} on the grid is a standard hexahedral finite element problem and any standard solvers can be employed.

The remaining question is how to choose $F_c^{\text{base}}$ in \eqref{eq:Ftrial-rotfree} without storing a physically meaningful deformation gradient. What we do have -- by construction of our particle stress unloading -- is an accurate resampling of Kirchhoff stress $\tau^n_c$ (extrinsically aggregated from particles) together with $V_{c}^n$. Because our materials are isotropic, the elastic energy and its tangent depend on the real $F$ only through its stretch $S$, where $F=RS$ is a polar decomposition; the previous right rotation is irrelevant to both energy and stiffness over a single backward‑Euler step. This motivates a rotation‑free stretch reference: we set
\begin{equation}
F_c^{\text{base}}:=S_c\quad\text{such that}\quad P(S_c)\,S_c^{\!\top}=\tau_c^n, \quad P(\cdot) = \nabla \psi(\cdot),
\label{eq:rotfree-def}
\end{equation}
i.e., we reconstruct only the stretch from the existing stress so that the center carries the same elastic state at the start of the step even if the (unknown) prior rotation is discarded. 

In \autoref{app:rotfree} we show that for isotropic $\psi$ this choice is objective and compared to keeping the rotation, the velocity discrepancy is only $O(\Delta t^2)$ per step.
\subsection{Stretch Reconstruction}
\label{sec:stretch-recon}

We assume an isotropic hyperelastic density that depends on $F$ only through its singular values.
Let $F=U\,\Sigma\,V^{\!\top}$ with $\Sigma=\operatorname{diag}(\sigma_1,\ldots,\sigma_d)$, $\sigma_i>0$, and write
\[
\psi(F)=\widehat\psi(\sigma_1,\ldots,\sigma_d)\quad\text{(symmetric in its arguments)}.
\]
For spectral energies, the principal values of the Kirchhoff stress $\tau=P(F)F^{\!\top}$ are
\begin{equation}
\label{eq:spectral-map}
\tau_i \;=\; \sigma_i\,\frac{\partial \widehat\psi}{\partial \sigma_i}(\sigma_1,\ldots,\sigma_d),
\qquad
\tau \;=\; U\,\operatorname{diag}(\tau_1,\ldots,\tau_d)\,U^{\!\top}.
\end{equation}
Thus given a symmetric $\tau$, the eigenvectors of $\tau$ provide the principal directions $U$, and the principal stretches $(\sigma_i)$ are recovered by solving the $d$ scalar equations in \eqref{eq:spectral-map}. 
A special case is when $\widehat\psi$ is strictly convex in the logarithmic stretches $e_i:=\log\sigma_i$, the map $e\mapsto (\tau_1,\ldots,\tau_d)$ is one-to-one. 
Once $(\sigma_i)$ are obtained, the rotation-free base is assembled as
\[
F^{\text{base}} = U\,\operatorname{diag}(\sigma_1,\ldots,\sigma_d)\,U^{\!\top},
\]
which realizes $P(F^{\text{base}})\,F^{\text{base}\,\top}=\tau$ by construction.

In computer graphics, MPM is best suited for simulating inelastic materials. A particularly useful $\psi$ is the St. Venant-Kirchhoff (StVK) model with Hencky strains, which enables simple plasticity return mapping for Drucker Prager sand \cite{klar2016sand}, (associative) Cam Clay snow \cite{gaume2018dynamic} and von Mises pasticine/metal \cite{gao2017agimp}. Another commonly seen choice is the deviatoric-dilational split Neo-Hookean model, proven successful for Herschel-Bulkley foam \cite{yue2015continuum} and (non-associative) Cam Clay fracture \cite{wolper2019cdmpm}. In fact, most of these elastoplastic materials (including \citet{stomakhin2013snow}'s snow) have extremely tiny elastic deformation before plasticity dominates, and the selection of $\psi$ is mostly for arriving at closed-form plastic return mapping rather than matching any nonlinear elastic behavior. Hence we derive the exact procedure of the $\tau \rightarrow \sigma_i$ mapping for these two models which covers the vast majority of scenarios where one would likely use MPM for.

\subsubsection{StVK with Hencky}

The model in terms of $(\sigma_i)$ and Lam\'e parameters $\mu,\lambda$ is
\begin{equation}
\label{eq:hencky-energy}
\psi(F)=\mu\,\sum_{i=1}^d (\log\sigma_i)^2 \;+\; \frac{\lambda}{2}\Big(\sum_{i=1}^d \log\sigma_i\Big)^2,
\end{equation}
whose first Piola derivative in spectral form is simple and 
\begin{equation}
P(F) = U (2\mu \Sigma^{-1} \log \Sigma + \lambda \operatorname{tr}(\log \Sigma) \Sigma^{-1}) V^T.
\end{equation}
The principal Kirchhoff stresses are 
\begin{equation}
\label{eq:hencky-taui}
\tau_i \;=\; \sigma_i\,\frac{\partial \widehat\psi}{\partial \sigma_i}
\;=\; 2\mu\,\log\sigma_i \;+\; \lambda\sum_{j=1}^d \log\sigma_j .
\end{equation}
Let $e_i:=\log\sigma_i$ and $s:=\sum_{j=1}^d e_j=\log J$ with $J=\prod_{j=1}^d \sigma_j$. 
Summing \eqref{eq:hencky-taui} over $i=1,\ldots,d$ gives
\begin{equation}
\label{eq:hencky-trace}
\sum_{i=1}^d \tau_i \;=\; (2\mu + d\lambda)\,s
\quad\Longrightarrow\quad
s \;=\; \frac{\operatorname{tr}\tau}{\,2\mu+d\lambda\,}.
\end{equation}
Then each logarithmic stretch follows directly:
\begin{equation}
\label{eq:hencky-ei}
e_i \;=\; \frac{\tau_i - \lambda s}{2\mu}
\;=\; \frac{\tau_i}{2\mu} \;-\; \frac{\lambda}{2\mu}\,\frac{\operatorname{tr}\tau}{\,2\mu+d\lambda\,}.
\end{equation}
Exponentiating yields the principal stretches
\begin{equation}
\label{eq:hencky-sigmai}
\sigma_i \;=\; \exp(e_i)
\;=\; \exp\!\left(\frac{\tau_i}{2\mu}\right)\,
\exp\!\left(-\frac{\lambda}{2\mu}\,\frac{\operatorname{tr}\tau}{\,2\mu+d\lambda\,}\right),
\end{equation}
which gives a global, closed-form inversion for all symmetric $\tau$ whenever $\mu>0$ and $\lambda>-2\mu/d$, and they ensure $J=\exp(s)>0$. 

\subsubsection{Split Neo–Hookean (deviatoric–volumetric form)}
\label{subsec:split-nh-correct}
Let $J=\det F$, $C=F^{\top}F$, $b=FF^{\top}$, and $\mathrm{dev}(b)=b-\tfrac{1}{d}\operatorname{tr}(b)\,I$. 
We adopt the standard isochoric/volumetric split by writing
\[
F = F^{\mathrm{dev}} F^{\mathrm{vol}},\qquad 
F^{\mathrm{dev}} = J^{-1/d}F,\qquad 
F^{\mathrm{vol}} = J^{1/d} I,
\]
and the split Neo–Hookean energy as
\begin{align}
\label{eq:split-nh-energy}
\Psi(F)\;&=\;\underbrace{\Psi^{\mu}\big(J^{-1/d}F\big)}_{\text{deviatoric}}
\;+\;
\underbrace{\Psi^{\kappa}(J)}_{\text{volumetric}},
\\
\Psi^{\mu}(F)&=\frac{\mu}{2}\big(\operatorname{tr}(F^{\top}F)-d\big),\\
\Psi^{\kappa}(J)&=\frac{\kappa}{2}\Big(\frac{J^{2}-1}{2}-\log J\Big),
\end{align}
where $\mu$ is the shear modulus and $\kappa$ is the bulk modulus ($\kappa=\tfrac{2}{3}\mu+\lambda$ in 3D, and $\kappa=\mu+\lambda$ in 2D). Equivalently, $\Psi^{\kappa}(J)=\tfrac{\kappa}{4}\,(J^{2}-1-2\log J)$. 
This replaces the naive (unsplit) $\tfrac{\mu}{2}(\operatorname{tr}C-d)$ by evaluating it on the \emph{isochoric} tensor $J^{-1/d}F$, so the first term depends only on $F^{\mathrm{dev}}$ while the second depends only on $J$ (i.e., $F^{\mathrm{vol}}$).  Differentiating \eqref{eq:split-nh-energy} yields the Kirchhoff stress in the familiar split form
\begin{align}
\tau(F)\;&=\;\mu\,J^{-2/d}\,\mathrm{dev}(b)\;+\;\alpha(J)\,I, \label{eq:split-nh-tau} \\
\alpha(J)\ :&=\ J\,\Psi^{\kappa}{}'(J)\ =\ \frac{\kappa}{2}\big(J^{2}-1\big).
\end{align}
Thus the deviatoric response is scaled by $J^{-2/d}$ and the volumetric response is purely spherical. 
Under isochoric motion ($J\equiv1$) we recover $\tau=\mu\,\mathrm{dev}(b)$; under pure dilation ($F=\gamma I$) we obtain $\tau=\alpha(J)I$ with $\alpha=\tfrac{\kappa}{2}(J^{2}-1)$. 

Let $\tau=U\,\mathrm{diag}(\tau_1,\dots,\tau_d)\,U^{\top}$ share eigenvectors with $b$, and write $\beta_i=\sigma_i^{2}$ where $F=U\,\mathrm{diag}(\sigma_1,\ldots,\sigma_d)\,V^{\top}$.  
From \eqref{eq:split-nh-tau},
\begin{equation}
\label{eq:split-nh-principal}
\tau_i\;=\;\mu\,J^{-2/d}\big(\beta_i-\bar\beta\big)\;+\;\alpha,\quad
\bar\beta:=\tfrac{1}{d}\sum_{j=1}^d \beta_j,\quad \alpha=\tfrac{\operatorname{tr}\tau}{d}.
\end{equation}
Hence the (spherical) part of $\tau$ directly determines $J$ via $\alpha(J)=\tfrac{\kappa}{2}(J^{2}-1)$,
\begin{equation}
\label{eq:J-from-alpha}
J\;=\;\sqrt{\,1+\tfrac{2}{\kappa}\,\alpha\,}\;=\;\sqrt{\,1+\tfrac{2}{\kappa}\,\tfrac{\operatorname{tr}\tau}{d}\,}\,,
\end{equation}
and the deviatoric part fixes the offsets 
\[
\delta_i\ :=\ \frac{\tau_i-\alpha}{\mu\,J^{-2/d}}\,,\quad \sum_i \delta_i=0.
\]
Let $m:=\bar\beta$. Then $\beta_i=m+\delta_i$ and the product constraint $\prod_i\beta_i=J^{2}$ gives a scalar equation for $m$:
\begin{align}
\text{(2D):}\quad & \beta_1\beta_2=(m+\delta)(m-\delta)=m^{2}-\delta^{2}=J^2 \\
&\ \Rightarrow\ 
m=\sqrt{J^2+\delta^{2}}\ \ (\delta:=\delta_1=-\delta_2), \label{eq:m-2D}
\end{align}
\begin{align}
\text{(3D):}\quad 
\prod_{i=1}^{3}(m+\delta_i) - J^2
&= m^{3}+m\,S_2+S_3 - J^2 = 0,  \label{eq:m-3D} \\ 
S_2:&=\!\!\sum_{i<j}\!\delta_i\delta_j,\quad S_3:=\delta_1\delta_2\delta_3.
\end{align}
We choose the unique real root with $\beta_i=m+\delta_i>0$; then $\sigma_i=\sqrt{\beta_i}$. In practice, \eqref{eq:m-2D} is closed form; \eqref{eq:m-3D} is a cubic and remains robust when the positive‑stretch branch is selected; see \autoref{sec:Cardano} for the detailed 3D Cardano solution.

\subsection{Material Mixture}

In cells interacting with multiple material particles we deliberately split kinematics from constitutive state. Mass and momentum are additive and the Eulerian velocity is single‑valued, so all species share velocities $v_c$ and $v_i$. In contrast, constitutive response is material‑specific. We therefore keep, at the same center locations, per‑material $k$ colocated quadrature $(V_{c,k},\tau_{c,k})$. They independently reconstruct stretches and contribute to the energy sum in \eqref{eq:incpot-rotfree}. Note that for explicit time integration there is no need for separate quadrature copies since the force \eqref{eqn:explicitforce} is additive with respect to $V \tau$.

\subsection{Water} 
\label{sec:water}

For inviscid fluids, particles do not possess shear strength and therefore do not require the storage of a full deformation gradient $F_p$. Instead, we only track the volumetric deformation $J_p = \det(F_p)$. To maintain consistency with our stress-transfer framework, we define a scalar Kirchhoff pressure $\pi$ derived from a volumetric strain energy density $\psi(J)$. 

We adopt a standard quadratic penalty model for the equation of state:
\begin{equation}
\psi(J) = \frac{\kappa}{2}(J-1)^2,
\end{equation}
where $\kappa$ is the bulk modulus. The corresponding scalar Kirchhoff stress (the hydrostatic component of $\tau$) is given by $\pi(J) = J \psi'(J) = \kappa J(J-1)$. Note that we use Kirchhoff stress rather than Cauchy stress to ensure the quantity scales correctly with the initial particle volume $V_p$ during transfer.

At the beginning of the time step, we evaluate the particle pressure $\pi_p^n = \pi(J_p^n)$ and unload the extensive pressure moment to the cell centers:
\begin{equation}
\pi_c^n = \frac{1}{V_c^n} \sum_p w_{cp} V_p \pi_p^n.
\end{equation}
Similar to the rotation-free stretch reconstruction for solids, we must reconstruct a base Jacobian $J_c^{\text{base}}$ that is consistent with the transferred pressure. Inverting the relation $\pi_c^n = \kappa J_c^{\text{base}} (J_c^{\text{base}} - 1)$ for positive $J$ yields:
\begin{equation}
J_c^{\text{base}} = \frac{1 + \sqrt{1 + 4\pi_c^n/\kappa}}{2}.
\end{equation}
With $J_c^{\text{base}}$ established, the implicit integration step minimizes the total potential energy. The trial Jacobian on the grid is defined as $J_c(v) = \det(I + \Delta t G_c(v)) J_c^{\text{base}}$, and the optimization becomes:
\begin{equation}
\min_{v} \sum_i \frac{1}{2}m_i \|v_i - v_i^n\|^2 + \sum_c V_c^n \psi(J_c(v)).
\end{equation}
After the grid solve, we compute the grid divergence $\nabla \cdot v_c^{n+1} = \operatorname{tr}(G_c^{n+1})$ or use the exact determinant to update the particle volume ratio:
\begin{equation}
J_p^{n+1} = \left( 1 + \Delta t \sum_c w_{cp} \operatorname{tr}(G_c^{n+1}) \right) J_p^n.
\end{equation}

\subsection{Degradation to FLIP/PIC Transfers}
\label{sec:flip-pic}
Although our scheme is designed with APIC in mind, it can be degraded to a classical FLIP/PIC-style particle-grid transfer when desired. We do not consider this FLIP/PIC variant a contribution of the paper, but include it for completeness. To obtain it, one simply omits the affine correction carried by the particle and center velocity gradient states when transferring momentum to the grid, so that particle-to-center and center-to-node transfers reduce to pure weighted averaging of velocities. For the transfer back to particles, one first interpolates both the updated nodal velocity and the nodal velocity increment over the time step to cell centers. The centered velocity increment is then interpolated to particles and added to the previous particle velocity to define a FLIP particle velocity, while the centered updated velocity is interpolated to particles to define a PIC particle velocity. The final particle velocity is formed by a linear blend between these FLIP and PIC velocities, and particle positions are advected using the PIC particle velocity.

\section{Algorithm}

\autoref{alg:driver} advances one MPM Lite time step by calling \emph{Unload} (\autoref{alg:unload}), \emph{Integrate} (\autoref{alg:grid-fem}) -- either explicit assembly or the incremental‑potential implicit solve that uses the stretch reconstruction in \autoref{sec:stretch-recon} -- and \emph{Load} (\autoref{alg:load}). 

The most important key to MPM Lite is that all steps only involve linear kernels, and \emph{Integrate} (\autoref{alg:grid-fem}) is pure FEM without accessing particles. The computational pattern greatly resembles FLIP/APIC fluids, where the integration is typically a finite difference/volume Poisson solver purely on the grid.

\algtext*{EndFunction}\algtext*{EndProcedure}\algtext*{EndFor}\algtext*{EndIf}\algtext*{EndWhile}
\algrenewcommand\algorithmicindent{0.8em}
\algrenewcommand\algorithmicrequire{\textbf{Input:}}
\algrenewcommand\algorithmicensure{\textbf{Output:}}

\begin{algorithm}[t]
\small
\caption{MPM Lite's One Time Step}
\label{alg:driver}
\begin{algorithmic}[1]
\Require $(x_p^{n},v_p^{n},F_p^{n},G_p^{n})$; materials $\{\psi_k\}$
\Ensure $(x_p^{n+1},v_p^{n+1},F_p^{n+1},G_p^{n+1})$
\State $(m_i,v_i^n,\{V_{c,k}^n,\tau_{c,k}^n\},v_c^n,G_c^n)\gets \Call{Unload}{x_p^{n},v_p^{n},F_p^{n},G_p^{n},\{\psi_k\}}$
\State $v_i^{n+1}\gets \Call{Integrate}{(m_i,v_i^n),\{V_{c,k}^n,\tau_{c,k}^n\},\{\psi_k\}}$
\State $(x_p^{n+1},v_p^{n+1},F_p^{n+1},G_p^{n+1})\gets \Call{Load}{v_i^{n+1}}$
\end{algorithmic}
\end{algorithm}

\begin{algorithm}[t]
\small
\caption{Unload Particle States}
\label{alg:unload}
\begin{algorithmic}[1]
\Require $(x_p^{n},v_p^{n},F_p^{n},G_p^{n},\{\psi_k\})$
\Ensure $(m_i,v_i^n,\{V_{c,k}^n,\tau_{c,k}^n\},v_c^n,G_c^n)$
\State \text{Clear all data on $i$ and $c$.}
\ForAll{$p$} 
  \State $\tau_p^n \gets \textsc{Stress}(F_p^n,\mathrm{mat}_p)$ \Comment{Kirchhoff stress}
  \State $k\gets \mathrm{mat}_p$  \Comment{Material ID, use $k=0$ for explicit integration}
  \ForAll{$c$ influenced by $p$}
    \State $m_c{+}{=}\,w_{cp}m_p$;\;$v_c^n{+}{=}\,w_{cp}m_p\!\big(v_p^n+G_p^n(x_c-x_p^n)\big)$;\; $G_c^n{+}{=}\,w_{cp}m_pG_p^n$
    \State $V_{c,k}^n{+}{=}\,w_{cp}V_p$;\; $\tau_{c,k}^n{+}{=}\,w_{cp}V_p\,\tau_p^n$
  \EndFor
\EndFor
\ForAll{$c$ with $m_c^n\neq 0$}
  \State $v_c^n\gets v_c^n/m_c^n$;\; $G_c^n\gets G_c^{n}/m_c^n$
  \ForAll{$k\in\mathcal{M}$ with $V_{c,k}^n \neq 0$} \Comment{$\mathcal{M} = \{0\}$ for explicit integration}
    \State $\tau_{c,k}^n\gets \tau_{c,k}^n/V_{c,k}^n$ 
  \EndFor 
\EndFor
\ForAll{$c$ with $m_c^n\neq 0$}
  \ForAll{$i\in\mathrm{Corners}(c)$}  \Comment{Gather; $w_{ic}\!=\!2^{-d}$}
    \State $m_i^n{+}{=}\,w_{ic} m_c^n$;\; $v_i^n{+}{=}\,w_{ic} m_c^n\!\big(v_c^n+G_c^n(x_i-x_c)\big)$
  \EndFor
\EndFor
\ForAll{$i$ with $m_i^n\neq 0$} \State $v_i^n\gets v_i^n/m_i^n$ \EndFor
\end{algorithmic}
\end{algorithm}

\begin{algorithm}[t]
\small
\caption{Integrate (Explicit or Implicit)}
\label{alg:grid-fem}
\begin{algorithmic}[1]
\Require $(m_i^n,v_i^n)$; $\{(V_{c,k}^n,\tau_{c,k}^n)\}$; $\{\psi_k\}$
\Ensure  $v_i^{n+1}$
\If{\textsc{Explicit}}  
  \State $f_i\gets -\sum_{c}\sum_{k} V_{c,k}^n\,\tau_{c,k}^n\,\nabla w_{ic}$ \Comment {$k=0$}
  \State $v_i^{n+1}\gets v_i^n+\Delta t f_i/m_i^n$
\Else \Comment{\textsc{Implicit}}
  \ForAll{$c$}
    \ForAll{$k$ with $V_{c,k}^n \neq 0$}
      \State Build $S_{c,k}^n$ from $\tau_{c,k}^n$ per \autoref{sec:stretch-recon}
    \EndFor
  \EndFor
  \State $v^{n+1} \gets \min_{v}\;\sum_{i}\tfrac12 m_i^n\|v_i\!-\!v_i^n\|^2+\sum_{c,k} V_{c,k}^n\,\psi_k\big((I+\Delta t\,G_c(v))S_{c,k}^n\big)$
\EndIf
\end{algorithmic}
\end{algorithm}

\begin{algorithm}[t]
\small
\caption{Load Information to Particles}
\label{alg:load}
\begin{algorithmic}[1]
\Require $v_i^{n+1}$
\Ensure $(x_p^{n+1},v_p^{n+1},F_p^{n+1},G_p^{n+1})$
\ForAll{$c$ with $m_c^n \neq 0$}
  \State $v_c^{n+1}\gets\sum_{i\in\mathrm{Corners}(c)} w_{ic}\,v_i^{n+1}$
  \State $G_c^{n+1}\gets\sum_{i\in\mathrm{Corners}(c)} v_i^{n+1}\otimes\nabla w_{ic}$
\EndFor
\ForAll{$p$}
  \State $v_p^{n+1}\gets\sum_c w_{cp}\,v_c^{n+1}$;\quad $G_p^{n+1}\gets\sum_c w_{cp}\,G_c^{n+1}$
  \State $x_p^{n+1}\gets x_p^n+\Delta t\,v_p^{n+1}$;\quad $F_p^{n+1}\gets (I+\Delta t\,G_p^{n+1})\,F_p^n$
\EndFor
\end{algorithmic}
\end{algorithm}

\section{Results}
This section evaluates the performance of our method across explicit and implicit settings, material versatility, and robustness to common simulation pathologies. In \S~\ref{subsec:explicit_comp} and \S~\ref{subsec:implicit_comp}, we compare our approach with existing methods under explicit and implicit formulations, respectively. Furthermore, in \S~\ref{subsec:vbd}, we demonstrate that our method can be readily coupled with off-the-shelf solvers such as VBD to achieve acceleration. In \S~\ref{subsec:versatil_material}, we show that MPM Lite can handle a wide range of common inelastic materials, including Cam-Clay fracture~\cite{wolper2019cdmpm}, Drucker–Prager sand~\cite{klar2016sand}, von Mises plasticity~\cite{li2022energetically}, snow~\cite{stomakhin2013snow}, and Herschel–Bulkley foam~\cite{yue2015continuum}. Finally, we evaluate the robustness of MPM Lite with respect to momentum conservation and plasticity optimization, and discuss the memory usage of our method in \S~\ref{subsec:angular_momentum}, \S~\ref{subsec:fixed_point}, and \S~\ref{subsec:memory}. We implemented MPM Lite and run all experiements on a workstation with an NVIDIA RTX Pro 6000 GPU and an Intel Core i9-9980XE CPU. All  code will be made publicly available.

\subsection{Explicit Comparision}
\label{subsec:explicit_comp}

In this subsection, we evaluate the performance of MPM Lite under explicit settings and compare it with traditional quadratic B-spline MPM and CK-MPM~\cite{liu2025ckmpm}. While the primary advantage of MPM Lite lies in its formulation for implicit integration, which completely eliminates the burden of grid-to-particle-to-grid transfers, we include explicit integration results here solely for completeness. Notably, MPM Lite also demonstrates improved performance in explicit settings compared to existing methods.

\begin{figure}
    \begin{minipage}{\linewidth}
        \centering
        \includegraphics[width=\linewidth]{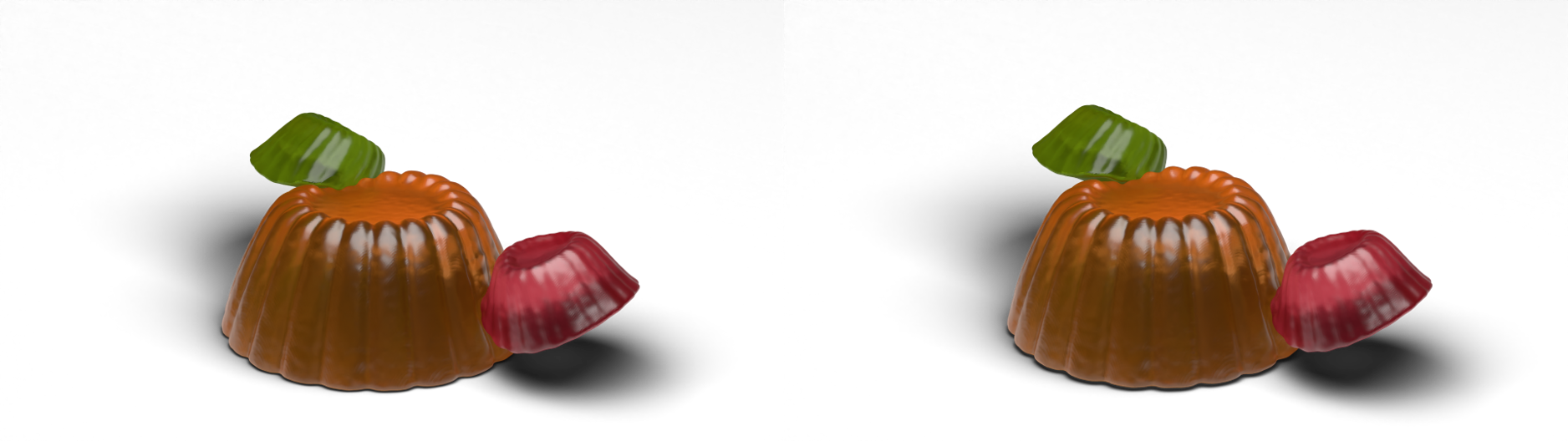}
    \end{minipage}
    \put(-1.0\linewidth+5,26){\scalebox{1}{\color{black} (a) CK-MPM}}
    \put(-0.5\linewidth+5,26){\scalebox{1}{\color{black} (b) Ours}}
    \caption{\textbf{Jelly Falling}. We simulate two jelly-like objects falling onto a third soft, elastic jelly using CK-MPM and our proposed MPM Lite, respectively.}
    \label{fig:jelly}
\end{figure}

\begin{table}[ht]
\caption{The total runtimes (in seconds) for the Jelly Falling example are reported. We compare the performance of the classic MPM scheme, CK-MPM~\cite{liu2025ckmpm}, and our proposed MPM Lite. All simulations run for 3 seconds at 120 frames per second, using a $\Delta x$ of $\frac{1}{256}$\,m.}
\label{tab:jelly}

\centering
\begin{tabular}{lcccc}
\hline
Method & MPM & CK-MPM & MPM Lite & Speedup \\
\hline
Jelly Falling & 404.1s & 247.7s & 215.4s & 1.88$\times$ \\
\hline
\end{tabular}

\end{table}

\textit{Jelly Falling}. We simulate two jelly-like objects falling onto a third soft, elastic jelly using CK-MPM and our proposed MPM Lite, respectively. All objects are modeled using a St. Venant–Kirchhoff (StVK) constitutive model with a Young’s modulus of $5\times10^{3}$\,Pa and a Poisson’s ratio of 0.4. This example consists of a total of 1.14M particles. We use a constant time step of $\Delta t = 6\times10^{-5}$\,s and run the simulation for 3 seconds at 120 frames per second, with a grid spacing of $\Delta x = \frac{1}{256}$\,m. Figure~\ref{fig:jelly} shows two visual snapshots captured at the moment of collision between the jelly objects. For quantitative comparison and performance evaluation, we report the total runtime in Table~\ref{tab:jelly}. By leveraging a linear kernel, our method employs a smaller stencil and enables faster particle-to-grid transfers, achieving a $1.88\times$ speedup over traditional quadratic B-spline explicit MPM, while also comparing favorably with CK-MPM.

\begin{figure}
    \begin{minipage}{\linewidth}
        \centering
        \includegraphics[width=\linewidth]{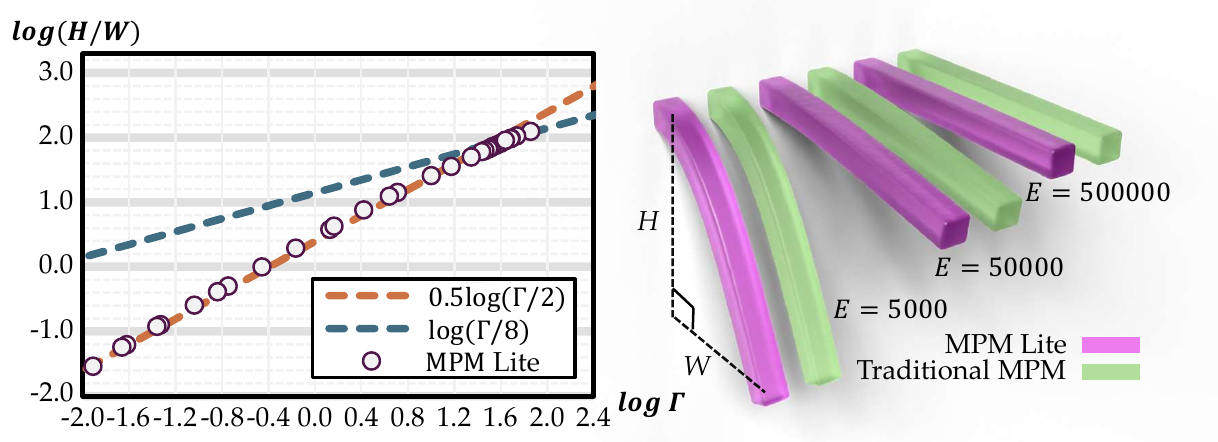}
    \end{minipage}
    \caption{\textbf{Cantilever Beams}. We present both quantitative and visual comparisons between MPM Lite and traditional implicit MPM. (Left) The elastic response produced by MPM Lite closely matches the theoretical predictions reported in~\cite{romero2021physical}. (Right) MPM Lite and traditional implicit MPM yield visually consistent deformation results for cantilever beams with varying stiffness.}
    \label{fig:cantilever}
\end{figure}

\begin{figure}
    \begin{minipage}{\linewidth}
        \centering
        \includegraphics[width=\linewidth]{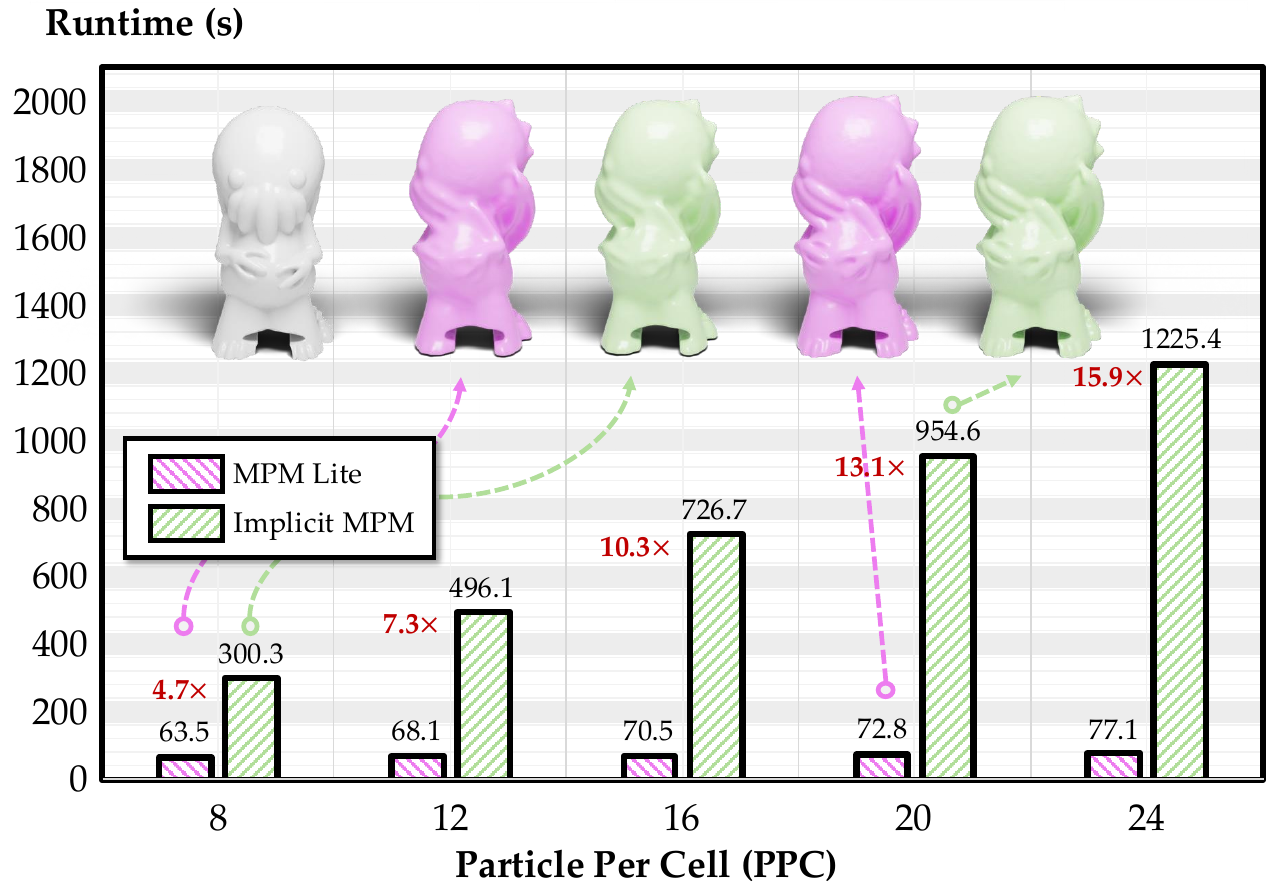}
    \end{minipage}
    \caption{\textbf{Speedup curve with respect to PPC}. A \textit{Faceless} object is twisted using MPM Lite and traditional implicit MPM under varying particles-per-cell (PPC) settings. The total runtime of each simulation is reported in the figure. MPM Lite achieves up to a $15.9\times$ speedup at 24 PPC. Notably, traditional MPM often requires relatively large PPC ($\ge 20$) to prevent numerical fracture.}
    \label{fig:ppc}
\end{figure}

\begin{figure*}
    \begin{minipage}{\textwidth}
        \centering
        \includegraphics[width=\textwidth]{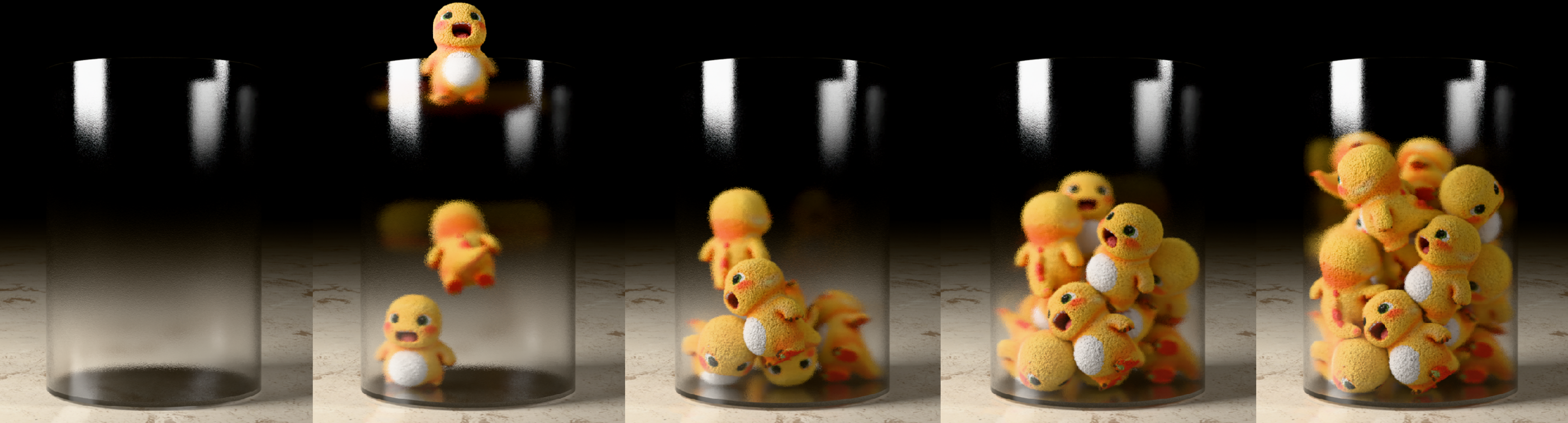}
    \end{minipage}
    \caption{\textbf{Stuffed Toys}. A total of 18 stuffed toys are dropped into a glass container. All toys share the same hyperelastic material model. The scene contains 5.22M particles in total, and MPM Lite coupled with VBD simulates the system at $0.22$s per time step.}
    \label{fig:nailong}
\end{figure*}

\begin{figure*}
    \begin{minipage}{\textwidth}
        \centering
        \includegraphics[width=\linewidth]{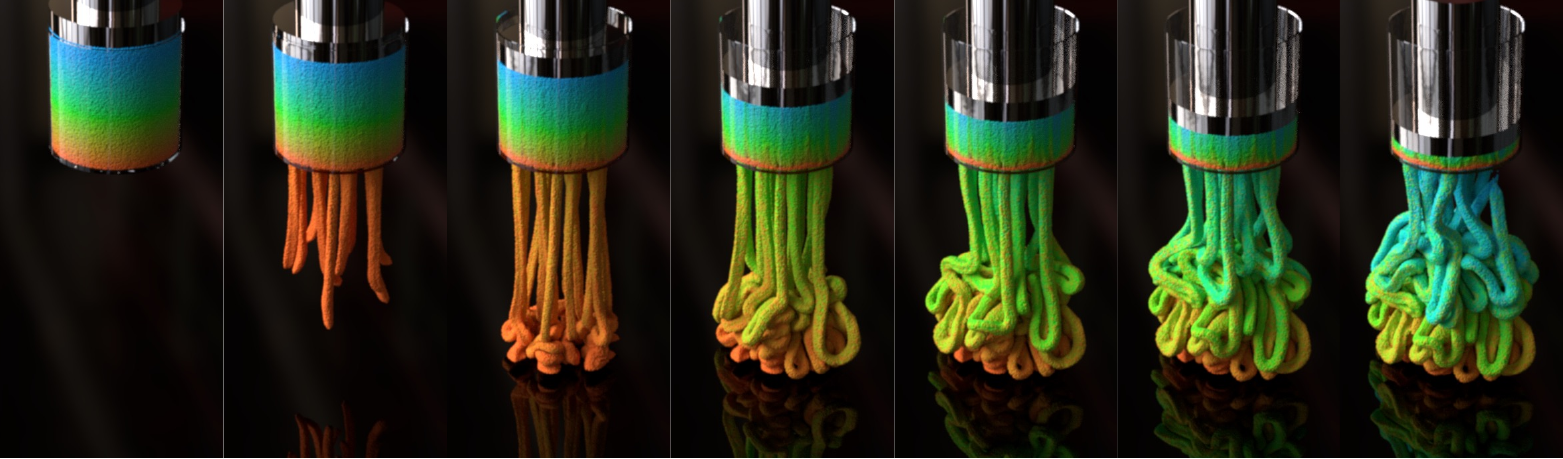}
    \end{minipage}
    \caption{\textbf{Noodles}. The noodle is undergoing large deformations as it is extruded through a cylindrical mold. The scene consists of in total 2.89M particles and the elasto-plastic behavior of noodle is modeled using von Mises plasticity.}
    \label{fig:noodles}
\end{figure*}

\subsection{Implicit Comparision}
\label{subsec:implicit_comp}

Here, we compare our method with traditional implicit MPM using quadratic B-spline basis functions. For a fair comparison, both methods employ the same preconditioned matrix-free conjugate gradient (PCG) solver for implicit integration.

\textit{Cantilever Beams}. We begin by validating the correctness of MPM Lite’s elastic response by comparing its results with those of classic implicit MPM as well as the theoretical solution for a cantilever beam. The right panel of Fig.~\ref{fig:cantilever} presents visual comparisons between MPM Lite and traditional implicit MPM under varying stiffnesses, specifically for three beams with Young’s moduli ranging from $5\times10^{3}$\,Pa to $5\times10^{5}$\,Pa. The results are visually consistent across all cases. To further verify our method, we compare our simulation results with theoretical predictions reported in~\cite{romero2021physical}. The aspect ratio ($H/W$, see the right panel of Fig.~\ref{fig:cantilever}) of a cantilever beam at static equilibrium is uniquely characterized as a function of the dimensionless gravito-bending parameter $\Gamma=\frac{12(1-\nu^2)\rho gL}{Eh^2}$, which is computed using the beam length ($L$), thickness ($h$), density ($\rho$), Young’s modulus ($E$), Poisson’s ratio ($\nu$), and gravitational acceleration ($g$). Our results converge to the red dashed line in the small-$\Gamma$ (high-stiffness) regime and to the blue dashed line in the large-$\Gamma$ (low-stiffness) regime, and smoothly follow the master curve throughout the transition between these two regimes.

\begin{figure}
    \begin{minipage}{\linewidth}
        \centering
        \includegraphics[width=\linewidth]{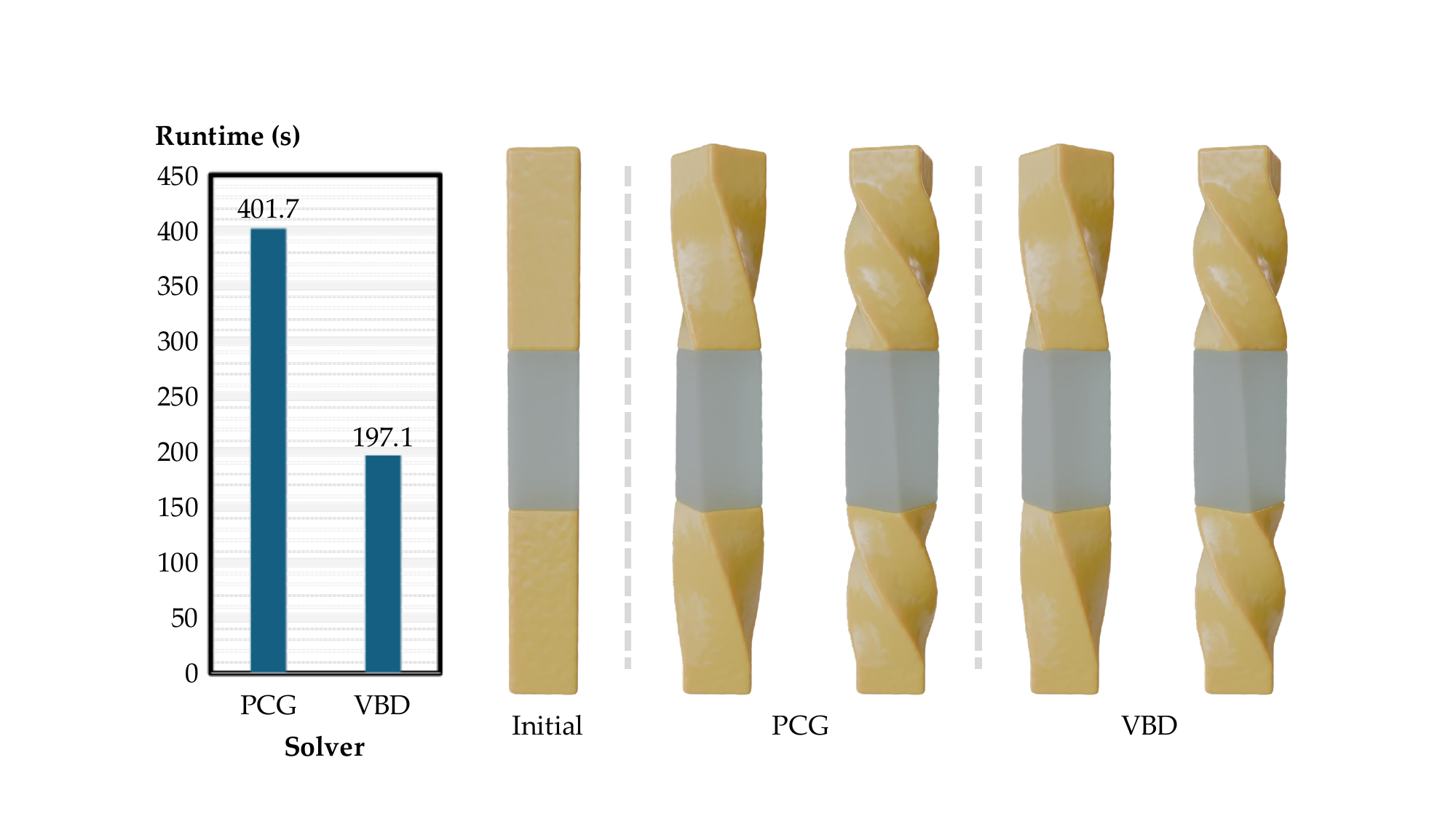}
    \end{minipage}
    \caption{\textbf{Speedup with VBD}. We simulate a twisting bar composed of two materials with different stiffnesses. We measure the total runtime of a 5-second simulation consisting of 500 frames at 100 frames per second, with a time step size of $\Delta t = 1\times10^{-3}$\,s. VBD achieves a $2\times$ speedup while maintaining visually similar behavior.}
    \label{fig:speedup_vbd}
\end{figure}

We then proceed to demonstrate the core advantage of MPM Lite: its implicit formulation, which is completely independent of particle states during the solve phase. By decoupling the implicit system from particle-dependent grid-to-particle and particle-to-grid transfers, MPM Lite enables a purely grid-based solve, substantially reducing computational overhead and improving efficiency. Importantly, this advantage becomes increasingly pronounced as the number of particles per cell (PPC) grows. High PPC is well known to be essential in MPM simulations to suppress numerical fracture and ensure stable material behavior, and practical simulations commonly require PPC values exceeding 20 for this reason. While traditional implicit MPM methods incur significantly higher computational costs as PPC increases, MPM Lite remains unaffected by PPC during the solve stage, making it particularly well suited for high-fidelity MPM simulations that demand large particle counts.

\textit{Speedup curve with respect to PPC}. We validate this claim in Fig.~\ref{fig:ppc} using a twisting example involving a faceless toy. In this experiment, the toy is twisted and then released, and we measure the total runtime for a 3-second simulation comprising 150 frames at 50 frames per second, with a time step size of $\Delta t = 1\times10^{-3}$\,s, under varying PPC values. The material is modeled with a Young’s modulus of $1\times10^{4}$\,Pa, a Poisson’s ratio of 0.3, and a density of $\rho = 1000\,$kg$\cdot$m$^{-3}$. We report speedups for PPC values ranging from 8 to 24. Across all cases, MPM Lite achieves significant speedups, reaching up to $15.9\times$ at 24 PPC.

\subsection{VBD}
\label{subsec:vbd}

As MPM Lite is highly modular and formulated as a hexahedral finite element–based incremental potential optimization, it can be readily coupled with off-the-shelf solvers such as PCG or multigrid PCG. In the following, we demonstrate how existing implicit solvers can benefit from our formulation. Specifically, we compare a Newton-PCG solver with the VBD~\cite{chen2024vertex} solver.

\begin{figure*}
    \begin{minipage}{\textwidth}
        \centering
        \includegraphics[width=\textwidth]{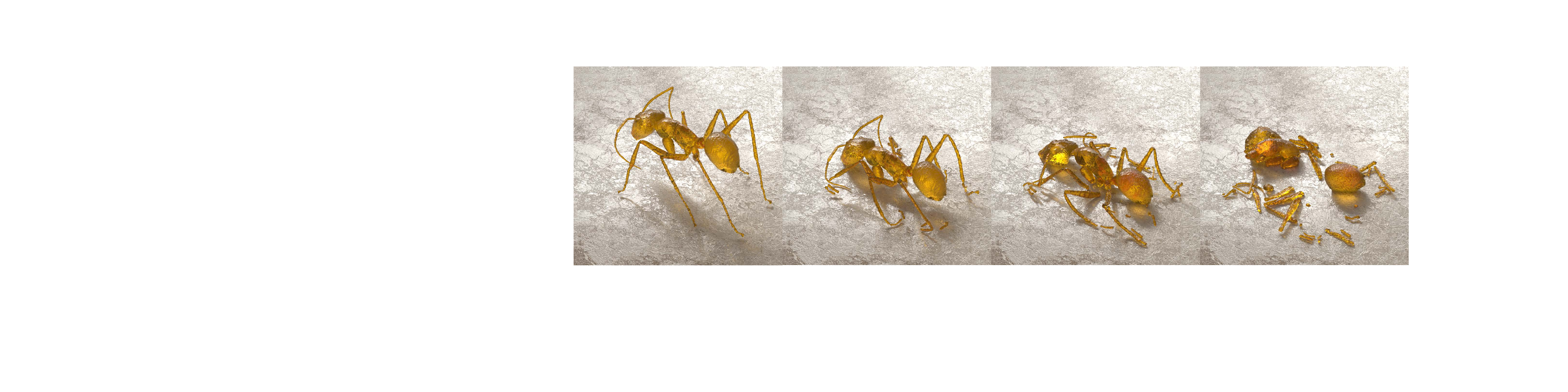}
    \end{minipage}
    \caption{\textbf{Candy Fracture}. We show the brittle fracture and fragmentation of a Camponotus ant-shaped candy upon impact. The simulation utilizes the Non-Associated Cam-Clay (NACC) model~\cite{wolper2019cdmpm} to capture the material’s breakup into multiple fragments. The scene includes 2.04M particles.}
    \label{fig:candy}
\end{figure*}

\begin{figure*}
    \begin{minipage}{\textwidth}
        \centering
\includegraphics[width=\textwidth]{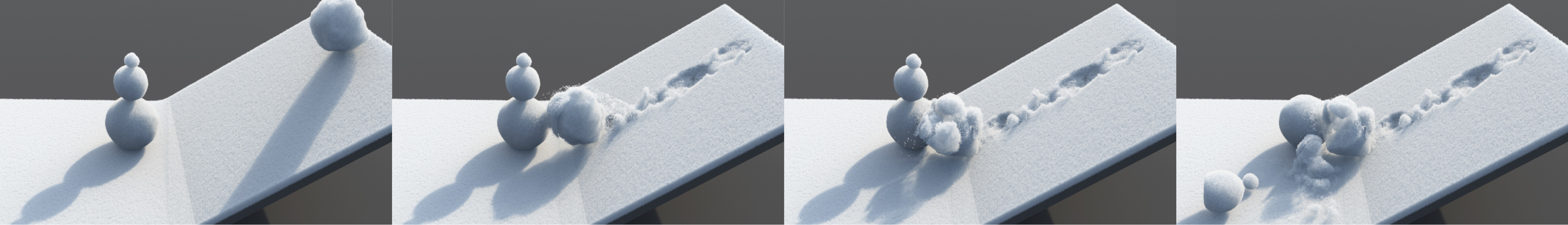}
    \end{minipage}
    \caption{\textbf{Rolling Snowball}. A snowball rolls down an inclined surface, accumulating surrounding snow and gradually increasing in size. Finally the snowball collides with a snowman at the end of the ramp and breaking into pieces. The scene contains 8.04M particles and snow in the scene is modeled using \cite{stomakhin2013snow}.}
    \label{fig:snow}
\end{figure*}
\begin{table*}[]
\centering
    \caption{\textbf{Parameters and Statistics.} We summarize the simulation parameters and timing statistics, including the maximum number of particles, the grid spacing $\Delta x$, the average runtime per frame in seconds, the integration scheme used in each simulation, the video frame time step size, the simulation time step size, the material models used for elasticity and plasticity, the density, Young's modulus, and Poisson's ratio. The abbreviations for material models are NACC for Non-Associated Cam-Clay~\cite{wolper2019cdmpm}, DP for Drucker--Prager~\cite{klar2016sand}, VM for von Mises~\cite{li2022energetically}, and HB for Herschel--Bulkley~\cite{yue2015continuum}. Material-related parameters are detailed in the last column as follows: 1) NACC: ($\xi,M,\beta,\alpha$); 2) DP: $\phi_f$; 3) VM: $\sigma_y$; 4) HB: ($h,\eta,\sigma_y$); and 5) snow: ($\theta_c,\theta_s$). In \textit{Sand and Water}, we use different densities for water and sand, and use $E$ in place of $\kappa$ for $J$-based water. In \textit{Rolling Snowball}, we use different snow materials for the snowball and the snowman.}
    \label{tab:params_stats}
\resizebox{\linewidth}{!}{
\begin{tabular}[t]{l r r r r r r r r r r r}
\toprule
Example
& Particle \#
& $\Delta x$ (m)
& s/frame
& Integration
& $\Delta t_{\text{frame}}$ (s)
& $\Delta t_{\text{step}}$ (s)
& Material
& $\rho$ (kg$\cdot$m$^{-3}$)
& $E$ (Pa)
& $\nu$
& Material Parameters
\\
\midrule

(Fig.~\ref{fig:nailong}) Stuffed Toys
& 5.23M & $4\times10^{-3}$ & 3.70 & VBD & 1/60 & $1\times10^{-3}$ & StVK
& $1\times10^3$ & $1\times10^4$ & $0.2$ & -\\

(Fig.~\ref{fig:noodles}) Noodles
& 2.89M & $8\times10^{-3}$ & 2.56 & PCG & $1/50$  & $1\times10^{-3}$ & VM
& $1\times10^3$ & $5\times10^6$ & $0.3$ & $9.6\times10^{3}$ \\

(Fig.~\ref{fig:candy}) Candy Fracture
& 2.04M & $1.5\times10^{-3}$ & 5.97 & PCG & $1/60$ & $1\times10^{-3}$ & NACC
& $2$ & $2\times10^4$ & $0.35$ & $(1.0, 2.36, 0.5, 0.953)$ \\

(Fig.~\ref{fig:snow}) Rolling Snowball
& 8.04M & $6.3\times10^{-3}$  & 14.43 & Explicit & 1/50 & $1\times10^{-4}$ & Snow & $\{3,1.2\}$ & $\{1,1.5\}\times10^3$ & $\{0.15, 0.2\}$ & $(0.01, 0.005)$ \\

(Fig.~\ref{fig:sand_water}) Sand and Water
& 1.93M & $5\times10^{-3}$ & 10.2 & PCG & $1/30$ & $1\times10^{-3}$ & DP
& $\{1,2\}\times10^3$ & $\{0.1,1\}\times10^6$ & $0.3$ & $30^\circ$\\

(Fig.~\ref{fig:wheel}) Wheel
& 1.23M & $5\times10^{-3}$ & 1.68 & PCG & $1/50$ & $1\times10^{-3}$ & VM
& $1\times10^3$ & $1\times10^8$ & 0.3 & $1.9\times10^{4}$ \\

(Fig.~\ref{fig:brownie}) Cream on Brownie
& 2.44M & $2\times10^{-3}$ & 2.83 & PCG & $1/60$ & $1\times10^{-3}$ & HB & $1.2\times10^3$
& $8.9\times10^3$ & $0.48$ & $(1.0, 16.0, 45)$\\

\bottomrule

\end{tabular}
}
\end{table*}

\textit{Speedup with VBD}. In Fig.~\ref{fig:speedup_vbd}, we consider a bar composed of two different materials. The two ends of the bar are assigned a Young’s modulus of $1\times10^{4}$\,Pa, while the middle section is assigned a Young’s modulus of $1\times10^{5}$\,Pa. In the experiment, one end of the bar is fixed, whereas the other end is gradually twisted. We measure the total runtime of a 5-second simulation consisting of 500 frames at 100 frames per second, with a time step size of $\Delta t = 1\times10^{-3}$\,s. The results show that VBD achieves a $2\times$ speedup while maintaining visually similar behavior, as illustrated in the right panel of Fig.~\ref{fig:speedup_vbd}.

\textit{Stuffed Toys}. To demonstrate the scalability of our method coupled with VBD, we present a more complex example simulated purely using VBD (see Fig.~\ref{fig:speedup_vbd}). In this experiment, we drop a total of 18 stuffed toys into a glass container. All stuffed toys share the same hyperelastic material model, with a Young’s modulus of $1\times10^{4}$\,Pa, a Poisson’s ratio of 0.2, and a density of $1\times10^{3}$\,kg$\cdot$m$^{-3}$. The toys collectively contain 5.22M particles, and MPM Lite coupled with VBD simulates the system at $0.22$\,s per time step.

\subsection{Versatile Materials}
\label{subsec:versatil_material}

\begin{figure*}
    \begin{minipage}{\textwidth}
        \centering
        \includegraphics[width=\textwidth]{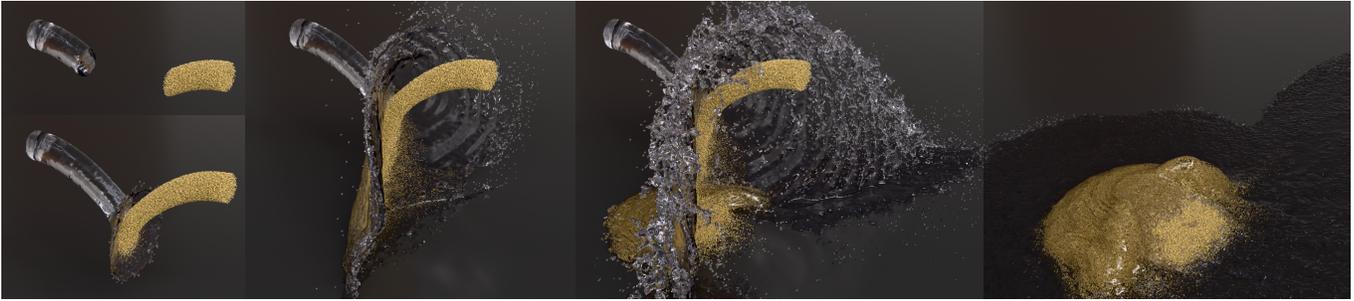}
    \end{minipage}
    \caption{\textbf{Sand and Water}. A coupled sand–water simulation under gravity, where continuous streams of water and sand are emitted from separate sources. There are 1.9M particles in the end of the simulation. As the water flows, it entrains and transports sand particles, producing splashing, erosion, mixing, and eventual deposition after the water disperses. Sand is modeled using Drucker-Prager plasticity~\cite{klar2016sand}, while water employs a $J$-based constitutive model ($\S$~\ref{sec:water}).}
    \label{fig:sand_water}
\end{figure*}

\begin{figure*}
    \begin{minipage}{\textwidth}
        \centering
    \includegraphics[width=\linewidth]{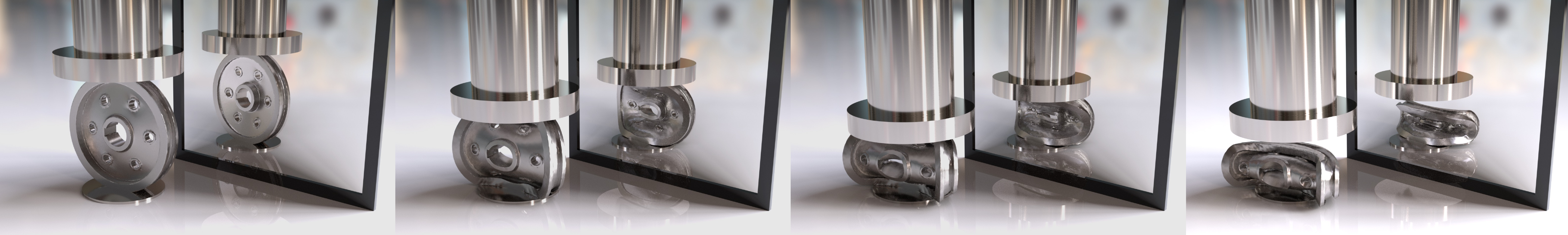}
    \end{minipage}
    \caption{\textbf{Wheel}. A stiff wheel is gradually pressed against a rigid surface under hydraulic loading, resulting in significant flattening and plastic deformation. The metal plasticity is modeled using the von Mises plasticity model. The simulation contains 1.23M particles.}
    \label{fig:wheel}
\end{figure*}

\begin{figure*}
    \begin{minipage}{\textwidth}
        \centering
        \includegraphics[width=\textwidth]{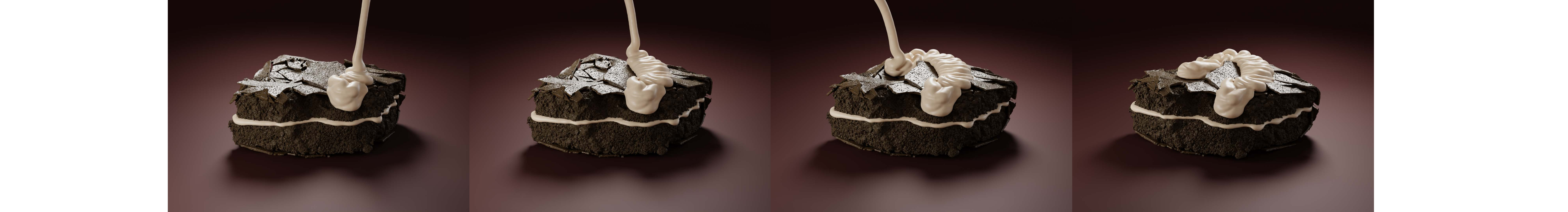}
    \end{minipage}
    \caption{\textbf{Cream on Brownie}. A visco-plastic cream is extruded onto a brownie and forms layered patterns on the surface. The cream is modeled using a Herschel–Bulkley material~\cite{yue2015continuum}, exhibiting yield stress and shear-dependent viscosity.}
    \label{fig:brownie}
\end{figure*}

In this part, we demonstrate the capability and scalability of MPM Lite. Versatile complex materials involving up to millions of particles can be efficiently simulated, including elasto-plastic noodles (Fig.~\ref{fig:noodles}), brittle fracture (Fig.~\ref{fig:candy}), snow (Fig.~\ref{fig:snow}), sand and water (Fig.~\ref{fig:sand_water}), metal (Fig.~\ref{fig:wheel}), and visco-plastic cream (Fig.~\ref{fig:brownie}). The timing statistics and material parameters for all scenes are summarized in Table~\ref{tab:params_stats}.

\textit{Noodles}. MPM Lite is capable of simulating elasto-plastic materials, such as noodles, with high fidelity. In Fig.~\ref{fig:noodles}, we demonstrate a noodle simulation modeled using von Mises plasticity, where the material undergoes large deformations as it is pressed through a cylindrical mold. This example highlights the ability of MPM Lite to robustly capture plastic flow behavior and complex shape changes while maintaining numerical stability during the extrusion process.

\textit{Candy Fracture}. As shown in Fig.~\ref{fig:candy}, a candy shaped like a Camponotus ant is dropped onto the ground and fractures into multiple pieces upon impact. To model the brittle fracture behavior in this example, we employ the Non-Associated Cam-Clay (NACC)~\cite{wolper2019cdmpm} model, which effectively captures the material’s failure and fragmentation characteristics.

\textit{Rolling Snowball}. In Fig.~\ref{fig:snow}, we simulate snow plasticity following the model of Stomakhin et al.~\cite{stomakhin2013snow}. A snowball is first formed and then released to roll down an inclined ramp, where it gradually accumulates snow from the surface. As the snowball rolls, it compresses and hardens the snow on the ramp while wrapping the accumulated material around itself, leading to continuous growth in size. Upon reaching the bottom of the ramp, the snowball crashes into a snowman standing at the end, causing the snowball to fragment and deform. This example demonstrates our method’s ability to capture snow compaction, hardening, accumulation, and impact-driven breakup.

\textit{Sand and Water}. Fig.~\ref{fig:sand_water} shows a coupled simulation of granular material and fluid interacting under gravity. In this example, two continuous streams of water and sand are emitted from separate sources. As the water flows, it entrains and transports sand particles, producing splashing, erosion, and mixing effects. The sand undergoes rearrangement and deposition while being carried by the fluid, eventually forming a mound after the water disperses. The sand is modeled using Drucker–Prager plasticity, while the water employs a $J$-based constitutive model as described in $\S$~\ref{sec:water}. This scenario highlights the capability of our framework to consistently simulate sustained multi-material flows and their evolving interactions under strong coupling.

\textit{Wheel}. Fig.~\ref{fig:wheel} show the capability of our method to robustly simulate hydraulic loading scenarios involving highly stiff material and large plastic deformations. In this experiment, hydraulic pressure is applied to gradually press the wheel against a rigid surface, leading to significant flattening and plastic deformation. A mirror placed beside the wheel provides an additional viewpoint to better visualize the deformation process. The wheel is modeled using von Mises plasticity with a high Young’s modulus of $1\times10^{8}$\,Pa. Despite the high stiffness of the aluminum material, the wheel exhibits pronounced plastic flow and contact-induced deformation while maintaining stable interaction with the rigid boundaries.

\begin{figure}
    \begin{minipage}{\linewidth}
        \centering
        \includegraphics[width=\linewidth]{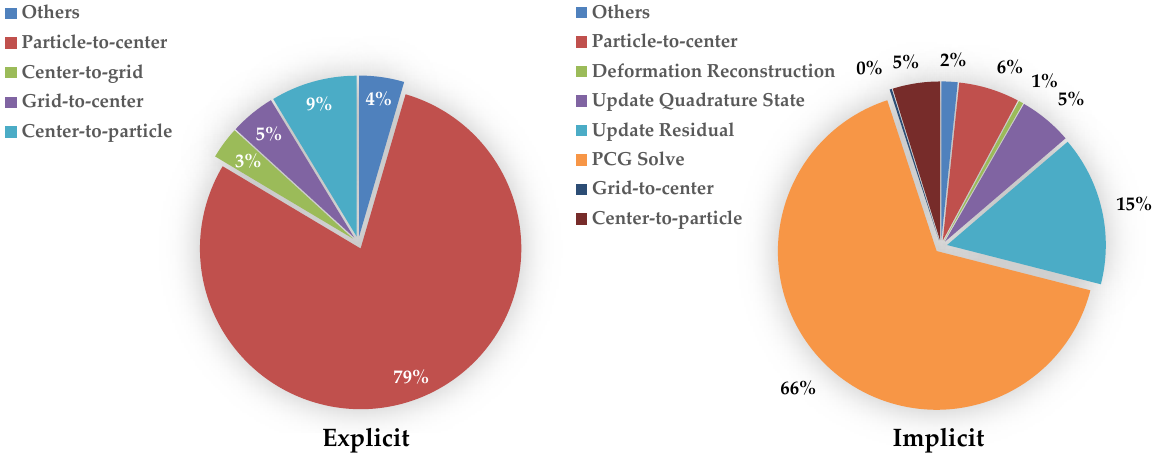}
    \end{minipage}
    \caption{\textbf{A typical breakdown of the total computational cost of our framework in explicit and implicit settings}. For explicit setting, we take the \textit{Jelly Falling} example for demonstration; for implicit setting, we take the \textit{Cream on Brownie} example for demonstration.}
    \label{fig:time_breakup}
\end{figure}

\textit{Cream on Brownie}. Fig.~\ref{fig:brownie} shows a visco-plastic cream being extruded onto a brownie. The cream is modeled using a Herschel–Bulkley non-Newtonian plastic material~\cite{yue2015continuum}, which exhibits a yield stress and shear-dependent viscosity. As the cream is dispensed, it forms a layered pattern on the brownie surface. Once deposited, the material gradually slows down and retains its shape due to its yield stress. Finally, we profile the total computational cost in Fig.~\ref{fig:time_breakup}.

\begin{figure}
    \begin{minipage}{\linewidth}
        \centering
        \includegraphics[width=\linewidth]{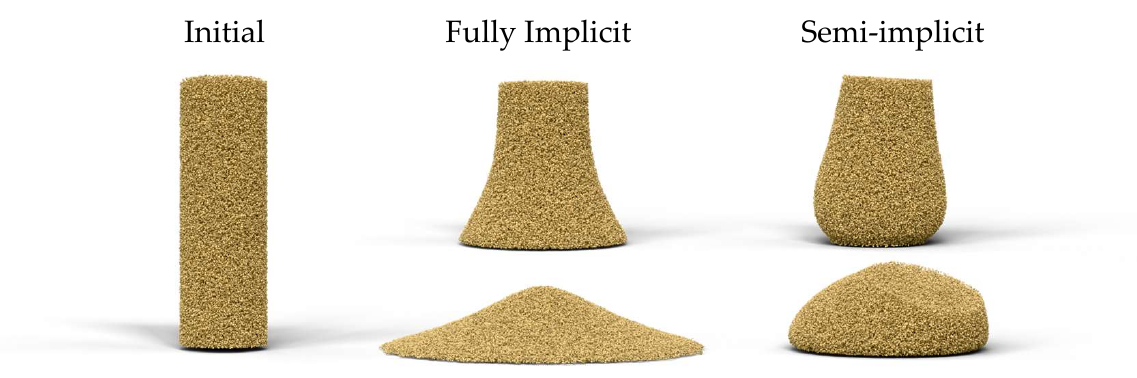}
    \end{minipage}
    \caption{\textbf{Fixed-point fully implicit plasticity and semi-implicit plasticity}. We compare the behavior of fixed-point fully implicit plasticity with that of semi-implicit plasticity. The fixed-point fully implicit approach correctly reconstructs the sand friction cone, whereas the semi-implicit scheme fails to produce accurate sand piling behavior.}
    \label{fig:fixed_point_plasticity}
\end{figure}

\subsection{Fixed-Point Plasticity}
\label{subsec:fixed_point}

Plasticity integration is commonly formulated as a nonlinear projection onto a yield surface, typically solved via return mapping as an independent procedure following elastic prediction. In the implicit formulation of MPM Lite, we instead employ a fully implicit fixed-point plasticity strategy. Specifically, plasticity is formulated as a fixed-point problem on the deformation gradient. Rather than treating plasticity as a standalone optimization, we update the plastic deformation within each Newton iteration, while the matrix-free conjugate gradient solve handles elasticity only.

Fig.~\ref{fig:fixed_point_plasticity} presents a comparison between the fully implicit scheme and a semi-implicit alternative. In the semi-implicit scheme, plastic return mapping is applied only after the Newton–PCG loop, which leads to nonphysical behavior such as the incorrect collapse of a sand pile. In contrast, MPM Lite coupled with fixed-point plasticity correctly captures the collapse into a stable friction cone, demonstrating the importance of tightly integrating plasticity within the implicit solver.

\subsection{Linear and Angular Momentum Study}
\label{subsec:angular_momentum}

Besides the momentum conservation derived theoretically, we also study the conservation of linear and angular momentum within the APIC framework of MPM Lite through two experiments.

\begin{figure}
    \begin{minipage}{\linewidth}
        \centering
        \includegraphics[width=\linewidth]{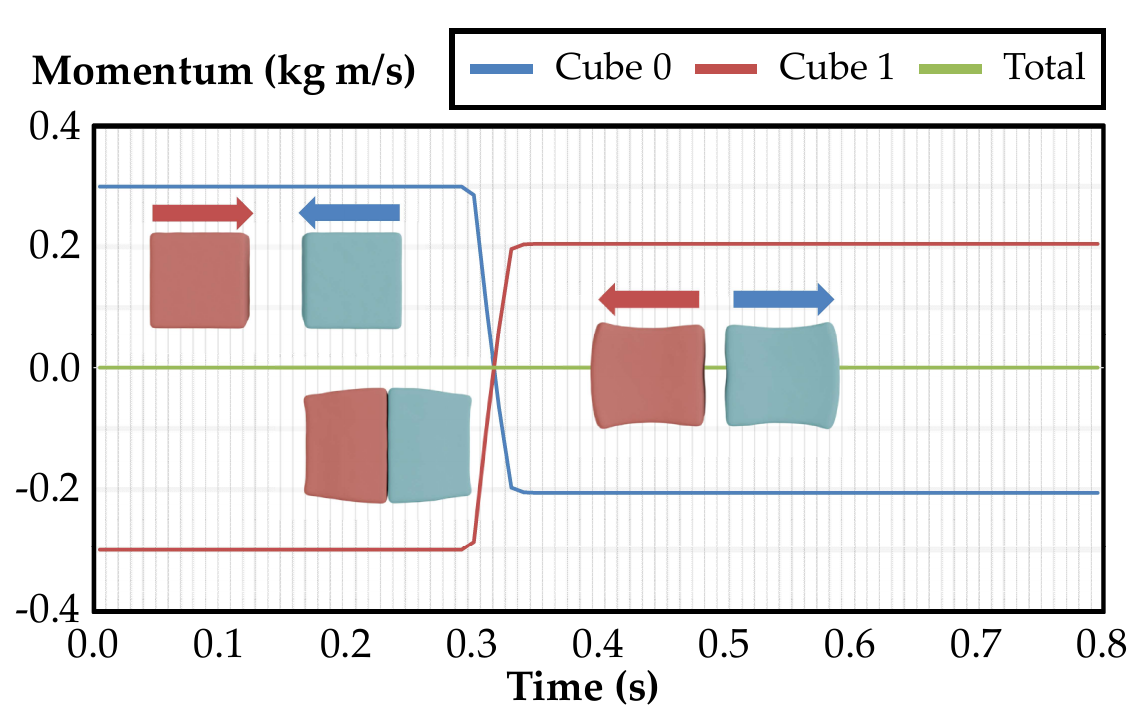}
    \end{minipage}
    \caption{\textbf{Conservation of Linear Momentum}. We study the conservation of linear momentum using two colliding elastic cubes. The total linear momentum of the system remains close to zero throughout the simulation.}
    \label{fig:linear_momentum}
\end{figure}

Below, we study the conservation of linear and angular momentum within the APIC framework of MPM Lite. The system’s linear momentum is conserved in the absence of external forces. To demonstrate this property, we conduct an experiment involving two elastic cubes of size $(0.1\mathrm{m},0.1\mathrm{m},0.1\mathrm{m})$ with identical material properties. The cubes collide with equal velocity magnitudes along the $x$-axis but in opposite directions. As shown in Fig.~\ref{fig:linear_momentum}, the total linear momentum remains consistently near zero throughout the simulation, with a maximum absolute value of $7.11\times10^{-15}\,\mathrm{kg}\cdot\mathrm{m}/\mathrm{s}$. The material parameters are set to a Young’s modulus of $5\times10^{3}\,\mathrm{Pa}$, a Poisson’s ratio of $0.3$, and a density of $1\times10^{3}\,\mathrm{kg}\cdot\mathrm{m}^{-3}$.

\begin{figure}
    \begin{minipage}{\linewidth}
        \centering
        \includegraphics[width=\linewidth]{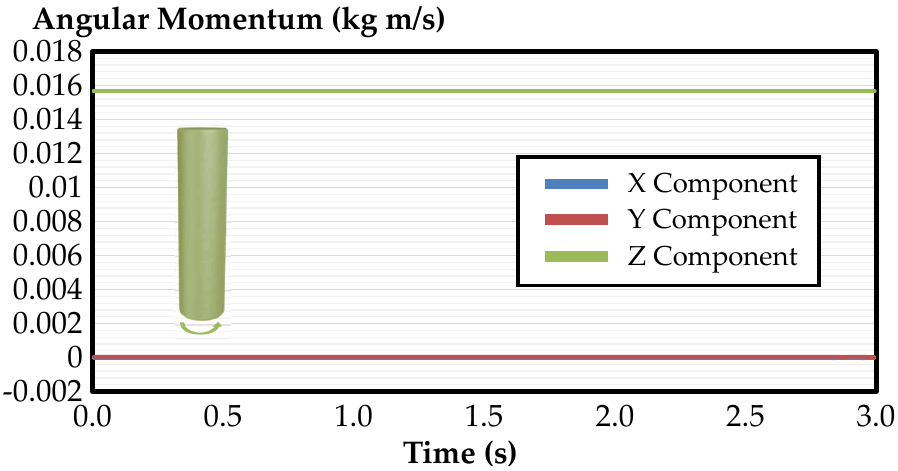}
    \end{minipage}
    \caption{\textbf{Conservation of Angular Momentum}. We study the conservation of angular momentum using a rotating rod. The $z$-component of the system’s angular momentum remains constant, while the other components remain zero throughout the simulation.}
    \label{fig:angular_momentum}
\end{figure}

The system’s angular momentum is also conserved, as shown in Fig.~\ref{fig:angular_momentum}. In this experiment, we initialize a rotating rod aligned with the $z$-axis, with a radius of $0.05\,\mathrm{m}$, a length of $0.4\,\mathrm{m}$, and an angular velocity of $(0,0,4\,\mathrm{rad/s})$. The $z$-component of angular momentum remains nearly constant throughout the simulation, with a relative error of $1.02\times10^{-4}$. The other components of angular momentum remain close to zero, with a maximum absolute value of $4.29\times10^{-5}\,\mathrm{kg}\cdot\mathrm{m}/\mathrm{s}$.

\subsection{Memory Usage}
\label{subsec:memory}

We discuss the memory usage of MPM Lite and compare them with traditional MPM formulations. In standard MPM, particles serve as moving quadrature points and actively participate in force assembly and time integration. As a result, both computation and memory usage scale with the number of particles, and in implicit settings, additional temporary buffers and solver-related data structures further amplify the memory footprint, especially at high PPC. In contrast, MPM Lite decouples particles from the integration process. Particles are used only to carry kinematic state and material history, while force assembly and time integration are performed entirely on fixed grid-aligned quadrature points and grid nodes. Consequently, the memory footprint of the solver depends primarily on the grid resolution rather than the number of particles. This design eliminates the need for particle-based integration buffers and avoids PPC-dependent growth in solver memory usage.

In explicit simulations, although MPM Lite requires additional memory to store center-based stress and deformation gradients, it achieves improved computational efficiency. This is similar to CK-MPM, which employs dual grids to store mass and momentum. As discussed in $\S$~\ref{sec:stress-transfer} and $\S$~\ref{sec:flip-pic}, the velocity gradient can be eliminated, and the symmetry of the stress tensor can be further exploited to reduce memory usage.

\section{Discussion}

\paragraph{Immunity to Hourglass Instabilities} A well-known pathology of one-point quadrature in hexahedral finite elements is the rank-deficiency of the stiffness matrix, which permits spurious ``hourglass'' modes (oscillatory nodal patterns that induce zero strain at the quadrature point). While these modes are theoretically admitted by the instantaneous grid solver, our kinematic transfer scheme prevents them from persisting or accumulating. The structural advantage of MPM Lite is that particle states ($x_p, F_p$) are not updated from grid nodes directly, but via the reconstructed cell-centered fields $v_c$ and $G_c$. We note that the standard trilinear hourglass modes lie strictly in the null space of the Node-to-Center projection operators; that is, the symmetric summation of an hourglass pattern yields exactly zero mean velocity and zero velocity gradient at the cell center. Consequently, the particles are kinematically decoupled from these spurious modes. Even if the grid velocity field $v_i^{n+1}$ transiently develops high-frequency hourglass noise, it is filtered out during the transfer and discarded with the grid at the end of the time step, ensuring long-term stability without the need for artificial stabilization forces.

\paragraph{Limitations}
Our rotation-free stretch  relies on isotropy: for anisotropic materials (e.g., fiber-reinforced or orthotropic models), the rotation carries material directions and cannot be discarded, so one must retain and transfer additional orientation state (or a full deformation representation) to recover the correct constitutive response. More broadly, our stretch reconstruction step depends on inverting a stress-to-stretch relation; while we provide robust procedures for common isotropic energies used in graphics, extending the framework to more general hyperelasticity or more complex inelastic models may require deriving new inverses, implementing iterative solves, and carefully handling ill-conditioning in extreme compression, extreme tension, or near-incompressible penalty settings. Because we integrate on a voxel mesh with a single cell-center quadrature, the method inherits the usual under-integration limitations of one-point hexahedral elements: although hourglass patterns are filtered by our transfer operators, accuracy can still degrade for bending-dominated motion, thin structures, or sharply varying stress fields, and additional quadrature or stabilization may be beneficial in such regimes. While the grid solve cost is decoupled from particles-per-cell, the overall pipeline still scales with particle count due to advection, resampling, and constitutive updates, and very sparse particle sampling can lead to noisy estimates near free surfaces or material interfaces. Finally, our current mixture treatment assumes a single shared velocity field per cell on a Cartesian background mesh; extending the approach to multi-velocity mixtures, richer coupling and contact models, and non-Cartesian or adaptive meshes remains future work.

\appendix

\section*{Appendix}

\section{Velocity Transfer Error Analysis} 
\label{appendix:v}

Here we provide a standalone derivation of the second-order accuracy for our two-hop transfers: P2G $v$, G2P $v$, and G2P $G$. By ``accuracy'' we measure the mismatch between our scheme and the standard quadratic B spline APIC transfers in traditional MPM.

\subsection{Notations and Transfer Schemes}
Grid nodes are indexed by $i$ with positions $x_i$, cell center quadratures by $c$ with positions $x_c$, and particles by $p$ with positions $x_p$. We use standard Q1 (multilinear) weights:
\[
\underbrace{w_{ic}}_{\text{center}\to\text{node}},\qquad
\underbrace{w_{cp}(x)}_{\text{particle position }x\to\text{center}},
\]
with the usual partition/first-moment identities:
\begin{align*}
&\sum_i w_{ic}=1,\ \ \sum_i w_{ic}x_i=x_c, \\
&\sum_c w_{cp}=1,\ \ \sum_c w_{cp}\,x_c=x
\quad(\text{for all }x).
\end{align*}
Let $\beta_{ip}:=\beta_{i}(x_p)$ denote the tensor-product quadratic $B_2$ spline weights (with $\sum_i\beta_{ip}=1$ and $\sum_i\beta_{ip}x_i=x$). 

Our two-hop P2G first accumulates at centers from particles
\begin{align*}
m_c&=\sum_p w_{cp}\,m_p,\quad \\
m_c v_c&=\sum_p w_{cp}\,m_p\Big(v_p+G_p(x_c-x_p)\Big),\quad \\
m_c G_c&=\sum_p w_{cp}\,m_p G_p,
\end{align*}
then scatter to nodes
\[
m_i=\sum_c w_{ic}\,m_c,\qquad
(mv)_i=\sum_c w_{ic}\,m_c\Big(v_c + G_c(x_i-x_c)\Big).
\]
The resulting nodal velocity is 
$$
v_i^{(2\text{hop})}=(mv)_i/m_i,
$$
while with B splines, the one-hop $B_2$ P2G velocity is
$$
v_i^{(\beta)} = \left(\sum_p \beta_{ip}\,m_p\Big(v_p + G_p(x_i-x_p)\Big)\right)/\left(\sum_p \beta_{ip} m_p \right).
$$

As for G2P, we have 
\begin{align*}
v_p^{(2\text{hop})}&=\sum_c w_{cp}\,v_c=\sum_c\sum_i w_{cp}w_{ic}\,v_i, \\
G_c&=\sum_i v_i\otimes\nabla w_{ic}\big|_{x_c},\qquad \\
G_p^{(2\text{hop})}&=\sum_c w_{cp}\,G_c=\sum_c\sum_i w_{cp}\,v_i\otimes\nabla w_{ic}\big|_{x_c},
\end{align*}
where with B splines, the one-hop B2 G2P with APIC is 
\begin{align*}
v_p^{(\beta)}&=\sum_i \beta_{ip}\,v_i \\
D_p:&=\sum_i \beta_{ip}\,(x_i-x_p)(x_i-x_p)^\top,\qquad\\
M_p:&=\sum_i \beta_{ip}\,v_i (x_i-x_p)^\top,\qquad\\
C_p^{(\beta)}&=M_p D_p^{-1},
\end{align*}
where $C_p$ is the notation commonly used in APIC for representing the MLS velocity gradient, and $D_p$ is a constant for quadratic/cubic B splines.

\subsection{Equality and Inequality Facts}
Both kernels have compact support of half-width $1.5 \Delta x$ per axis, hence for any contributing pair $(i,p)$ in dimension $d \in \{1,2,3\}$,
\begin{align}
\|x_i-x_p\|^k\ &\le\ 1.5^k\sqrt{d}^k\,\Delta x^k,\quad k\in\{1,2,3\}, \label{eq:stencil-geometry}
\end{align}
giving us useful bounds on powers of $\|x_i-x_p\|$. 

We also have some common reproduction identities. From the Q1/$B_2$ properties we will use:
\begin{align*}
&\sum_i w_{ic}=1,\ \ \sum_i w_{ic}x_i=x_c,\ \ \\
&\sum_i\nabla w_{ic}\big|_{x_c}=0,\ \ \sum_i (x_i-x_c)\otimes\nabla w_{ic}\big|_{x_c}=I,\\
&\sum_c w_{cp}=1,\ \ \sum_c w_{cp}x_c=x_p,\qquad \\
&\sum_i \beta_{ip}=1,\ \ \sum_i \beta_{ip}x_i=x_p.
\end{align*}
Moreover, in 1D one checks directly that for the same particle-node offset $s=(x_i-x_p)/\Delta x$,
the quadratic $B_2$ weight is bounded by the two-hop (Q1$\to$center, Q1$\to$node) effective weight,
\[
\beta_{1\mathrm{D}}(s)\le 1.5\ \Big(\sum_c w_{cp}w_{ic}\Big)_{1\mathrm{D}}(s),
\]
and by tensor products this yields, in $d$D,
\begin{equation}\label{eq:beta-le-twohop}
\beta_{ip}\ \le\ (1.5)^d\ \sum_c w_{cp}\,w_{ic}, \quad \forall (i,p).
\end{equation}

\subsection{Assumptions}
We will assume a \emph{minimal local smoothness} on one stencil. Specifically, we assume the Eulerian velocity $v:\mathbb{R}^d\rightarrow\mathbb{R}^d$ satisfies, on the small patch that influences the node/particle under consideration:
\begin{itemize}
  \item $v$ has bounded second derivatives: $\|H_v\|_{\infty,\text{patch}}<\infty$ (here $H_v$ is the componentwise Hessian);
  \item the Hessian is locally Lipschitz: there exists $L_{\text{Hess}}$ such that $\|H_v(x)-H_v(y)\|\le L_{\text{Hess}}\|x-y\|$ on the patch.
\end{itemize}
This is a ``$C^{2,1}$'' (Lipschitz Hessian) assumption.
It is natural in MPM/APIC because we apply it \emph{locally} on a single compact stencil (where the field is smooth between shocks/contacts),
so $\|H_v\|_{\infty,\text{patch}}$ and $L_{\text{Hess}}$ are finite by compactness.

\subsection{Error Bounds}

\subsubsection{P2G Velocity}
 
\begin{theorem}[P2G $v$ is second-order accurate]
\label{thm:p2g-v-twohop}
Let $i$ be a grid node of spacing $\Delta x$ and let particles $p$ carry $(m_p,v_p,G_p,x_p)$.
Define the two weight families
\[
\alpha_p := m_p \sum_{c} w_{cp}\,w_{ic},
\qquad
\beta_p  := m_p\,\beta_{ip},
\]
with normalizations $m_i^{(2\mathrm{hop})}:=\sum_p \alpha_p$ and $m_i^{(\beta)}:=\sum_p \beta_p$.
Set
\begin{align*}
v_p(i)&:= v_p + G_p\,(x_i-x_p),
\\
v_i^{(2\mathrm{hop})}&:=\frac{1}{m_i^{(2\mathrm{hop})}}\sum_p \alpha_p\,v_p(i),
\\
v_i^{(\beta)}&:=\frac{1}{m_i^{(\beta)}}\sum_p \beta_p\,v_p(i).
\end{align*}
Assume the continuum velocity $v$ is $C^{2,1}$ on the compact stencil that contributes to node $i$
(i.e.\ the componentwise Hessian is bounded and Lipschitz there) and that the support geometry
satisfies $\|x_i-x_p\|\le 1.5\sqrt{d}\,\Delta x$ for all contributing $(i,p)$.
Then
\begin{equation*}
\boxed{
\big\|v_i^{(2\mathrm{hop})}-v_i^{(\beta)}\big\|
\;\le\;
\underbrace{\big(1+(1.5)^d\big)\,1.5^2\,d}_{=:C(d)}\;
\|H_v\|_{\infty,\text{patch}_i}\;\Delta x^2 .
}
\end{equation*} 
Consequently the discrepancy is $O(\Delta x^2)$ with a constant that is independent of occupancy.
\begin{proof}

\emph{Step 1.} We start with a simple algebraic identity.
For any two weighted averages of the same list $\{\theta_p\}$,
$A=\frac{\sum\alpha_p \theta_p}{\sum\alpha_p}$ and $B=\frac{\sum\beta_p \theta_p}{\sum\beta_p}$,
\begin{equation}\label{eq:subtract-mean}
A-B
=\frac{1}{\sum\alpha_p}\sum_p (\alpha_p-\beta_p)\,(\theta_p-B).
\end{equation}
Indeed,
$\sum\alpha_p(\theta_p-B)=\sum\alpha_p \theta_p-B\sum\alpha_p
=\sum\alpha_p \theta_p-\frac{\sum\alpha_p}{\sum\beta_p}\sum\beta_p \theta_p$,
and dividing by $\sum\alpha_p$ yields \eqref{eq:subtract-mean}.
Apply \eqref{eq:subtract-mean} with $\theta_p=v_p(i)$, $A=v_i^{(2\mathrm{hop})}$ and $B=v_i^{(\beta)}$:
\begin{equation}\label{eq:key-identity}
v_i^{(2\mathrm{hop})}-v_i^{(\beta)}
=\frac{1}{m_i^{(2\mathrm{hop})}}
\sum_p\big(\alpha_p-\beta_p\big)\,\Big(v_p(i)-v_i^{(\beta)}\Big).
\end{equation}

\medskip
\emph{Step 2.} Next we study the deviations $v_p(i)-v_i^{(\beta)}$.
Let $x\mapsto v(x)$ be the underlying smooth field.
Fix particle $p$ and expand $v$ at $x_p$ towards $x_i$:
\begin{align*}
v(x_i)&=v(x_p)+\nabla v(x_p)(x_i-x_p)+R_{p\to i},
\\
\|R_{p\to i}\|&\le \tfrac12\,\|H_v\|_{\infty,\text{patch}_i}\,\|x_i-x_p\|^2 .
\end{align*}
By definition $v_p(i)=v_p+G_p(x_i-x_p)$,
and for affine data ($v_p=v(x_p)$, $G_p=\nabla v(x_p)$) we have $v_p(i)=v(x_i)$ exactly.
Thus in the general smooth case
\begin{align}
v_p(i)-v(x_i)&=-\,R_{p\to i}, \nonumber
\\
\Rightarrow \|v_p(i)-v(x_i)\|&\le \tfrac12\,\|H_v\|_{\infty,\text{patch}_i}\,\|x_i-x_p\|^2 .\label{eq:ap-vxi}
\end{align}
Because $v_i^{(\beta)}$ is a \emph{convex} (nonnegative, normalized) linear combination of $\{v_p(i)\}_p$,
\begin{align}
\|\,v_i^{(\beta)}-v(x_i)\,\| \;&\le\; \max_p \|\,v_p(i)-v(x_i)\,\|  \nonumber \\
\;&\le\; \tfrac12\,\|H_v\|_{\infty,\text{patch}_i}\,\max_p\|x_i-x_p\|^2 . \label{eq:beta-2nd-order}
\end{align}
Combining \eqref{eq:ap-vxi}-\eqref{eq:beta-2nd-order} gives for every $p$
\begin{align}
\|\,v_p(i)-v_i^{(\beta)}\,\|
\;&\le\; \|v_p(i)-v(x_i)\| + \|v_i^{(\beta)}-v(x_i)\| \nonumber\\
\;&\le\; \|H_v\|_{\infty,\text{patch}_i}\,\max_p\|x_i-x_p\|^2 . \label{eq:dev-bound}
\end{align}
By the compact support of the stencils,
$\|x_i-x_p\|\le 1.5\sqrt d\, \Delta x$, hence
\begin{equation}\label{eq:dev-bound-geom}
\|\,v_p(i)-v_i^{(\beta)}\,\|\ \le\ 1.5^2\,d\,\|H_v\|_{\infty,\text{patch}_i}\;\Delta x^2
\end{equation}
uniformly over all contributing particles $p$.

\medskip
\emph{Step 3.} Finally lets bound the kernel differences. Start from \eqref{eq:key-identity} and apply the triangle inequality:
\[
\big\|v_i^{(2\mathrm{hop})}-v_i^{(\beta)}\big\|
\ \le\ \frac{1}{m_i^{(2\mathrm{hop})}}
\Big(\max_p \|v_p(i)-v_i^{(\beta)}\|\Big)\;
\sum_p |\alpha_p-\beta_p| .
\]
For the sum of weight differences, use
$|\alpha_p-\beta_p|\le \alpha_p+\beta_p$ and sum over $p$:
\begin{equation}\label{eq:sum-diff}
\sum_p |\alpha_p-\beta_p|\ \le\ m_i^{(2\mathrm{hop})}+m_i^{(\beta)} .
\end{equation}
Finally use \eqref{eq:beta-le-twohop},
\begin{equation}\label{eq:beta-vs-twohop}
\beta_{ip}\ \le\ (1.5)^d\sum_c w_{cp}\,w_{ic}
\quad\Longrightarrow\quad
m_i^{(\beta)}\ \le\ (1.5)^d\,m_i^{(2\mathrm{hop})}.
\end{equation}
Insert \eqref{eq:dev-bound-geom}, \eqref{eq:sum-diff}, and \eqref{eq:beta-vs-twohop}:
\begin{align*}
&\big\|v_i^{(2\mathrm{hop})}-v_i^{(\beta)}\big\| \\
\ &\le\
\frac{1}{m_i^{(2\mathrm{hop})}}\;
\Big(1.5^2\,d\,\|H_v\|_{\infty,\text{patch}_i}\,\Delta x^2\Big)\;
\Big(m_i^{(2\mathrm{hop})}+m_i^{(\beta)}\Big) \\
\ &\le\
\big(1+(1.5)^d\big)\,1.5^2\,d\;\|H_v\|_{\infty,\text{patch}_i}\,\Delta x^2 .
\end{align*}
The factor $m_i^{(2\mathrm{hop})}$ cancels, i.e., the bound is occupancy-free.
\end{proof}
\end{theorem}

\subsubsection{G2P Velocity}

\begin{theorem}[G2P $v$ is second order accurate]
\label{thm:g2p-v-twohop-verbose}
Fix a particle $p$ at $x_p$. Let
\begin{align*}
T_{ip}&:=\sum_{c} w_{cp}\,w_{ic},\\
v_p^{(2\mathrm{hop})}&:=\sum_i T_{ip}\,v(x_i),
\\
v_p^{(\beta)}&:=\sum_i \beta_{ip}\,v(x_i).
\end{align*}
Assume $v$ is $C^{2,1}$ on the compact stencil that influences $p$. Then 
\[
\boxed{\quad
\|\,v_p^{(2\mathrm{hop})}-v_p^{(\beta)}\,\|
\ \le\ 1.5^2 d \ \|H_v\|_{\infty,\text{patch}_p}\ \Delta x^2.}
\]

\begin{proof}
\emph{Step 1.} 
Both families $\{T_{ip}\}_i$ and $\{\beta_{ip}\}_i$ are nonnegative and satisfy
$\sum_i T_{ip}=\sum_i \beta_{ip}=1$, $\sum_i T_{ip} x_i=\sum_i \beta_{ip} x_i=x_p$.
Hence
\begin{equation}\label{eq:recenter}
v_p^{(2\mathrm{hop})}-v_p^{(\beta)}
=\sum_i (T_{ip}-\beta_{ip})\Big(v(x_i)-v_p^{(\beta)}\Big),
\end{equation}
which is the difference-of-averages identity with the common mean subtracted.

\medskip
\emph{Step 2.} 
For each contributing node $i$,
\begin{align}
v(x_i)&=v(x_p)+\nabla v(x_p)(x_i-x_p)+\tfrac12 H_v(x_p):(x_i-x_p)^{\otimes2}+R_i, \nonumber \\
\|R_i\|&\le \tfrac16 L_{\mathrm{Hess}}\|x_i-x_p\|^3. \label{eq:taylor}
\end{align}
Averaging \eqref{eq:taylor} with $\beta_{ip}$ and using $\sum_i \beta_{ip}=1$, $\sum_i \beta_{ip}(x_i-x_p)=0$ gives
\begin{align*}
v_p^{(\beta)}&=v(x_p)+\tfrac12 H_v(x_p):D_p^{(\beta)} + R^{(\beta)},\\
D_p^{(\beta)}:&=\sum_i \beta_{ip}(x_i-x_p)^{\otimes2},\quad
R^{(\beta)}:=\sum_i \beta_{ip} R_i .
\end{align*}
Subtracting these two expressions yields, for each $i$,
\begin{align}
v(x_i)-v_p^{(\beta)}
&=\nabla v(x_p)(x_i-x_p)
+\tfrac12 H_v(x_p):\Big((x_i-x_p)^{\otimes2}-M_p^{(\beta)}\Big) \nonumber\\
&+\big(R_i - R^{(\beta)}\big).\label{eq:node-dev}
\end{align}

\medskip
\emph{Step 3.} Insert \eqref{eq:node-dev} into \eqref{eq:recenter} and cancel constant/linear parts.
Multiplying \eqref{eq:node-dev} by $(T_{ip}-\beta_{ip})$ and summing over $i$:
\begin{align*}
&\sum_i (T_{ip}-\beta_{ip})\,\nabla v(x_p)(x_i-x_p)\\
&=\nabla v(x_p)\Big(\underbrace{\sum_i T_{ip}(x_i-x_p)}_{0}-\underbrace{\sum_i \beta_{ip}(x_i-x_p)}_{0}\Big)=0,
\end{align*}
and
\begin{align*}
&\sum_i (T_{ip}-\beta_{ip})\,\Big(\tfrac12 H_v(x_p):D_p^{(\beta)}\Big) \\
&=\tfrac12 H_v(x_p):D_p^{(\beta)} \underbrace{\sum_i (T_{ip}-\beta_{ip})}_{0}=0.
\end{align*}
Therefore
\begin{align}
&v_p^{(2\mathrm{hop})}-v_p^{(\beta)} \nonumber \\
&=\tfrac12\,H_v(x_p):\Big(D_p^{(T)}-D_p^{(\beta)}\Big)
\ +\ \sum_i (T_{ip}-\beta_{ip})\,R_i, \label{eq:weighted-identity}
\end{align}
where $D_p^{(T)}:=\sum_i T_{ip}(x_i-x_p)^{\otimes2}$. 

\medskip
\emph{Step 3.} Finally let's bound the two terms in \eqref{eq:weighted-identity}. Both $D_p^{(T)}$ and $D_p^{(\beta)}$ are positive semidefinite convex averages of $(x_i-x_p)(x_i-x_p)^\top$, hence using \eqref{eq:stencil-geometry} we have
\begin{align*}
\|D_p^{(T)}\| &\le \operatorname{tr}D_p^{(T)}=\sum_i T_{ip}\|x_i-x_p\|^2\le 1.5^2 d\,\Delta x^2,
\\
\|D_p^{(\beta)}\| &\le 1.5^2 d\,\Delta x^2.
\end{align*}
Thus
\begin{align*}
&\Big\|\tfrac12\,H_v(x_p):\big(D_p^{(T)}-D_p^{(\beta)}\big)\Big\| \\
\ &\le\ \tfrac12\,\|H_v\|_{\infty,\text{patch}_p}\,\big(\|D_p^{(T)}\|+\|D_p^{(\beta)}\|\big) \\
\ &\le\ 1.5^2 d\,\|H_v\|_{\infty,\text{patch}_p}\,\Delta x^2 .
\end{align*}
The remainder term can also be easily bounded. 
From \eqref{eq:taylor} and \eqref{eq:stencil-geometry},
$\|R_i\|\le \tfrac16 L_{\mathrm{Hess}}\,1.5^3 d^{3/2} \Delta x^3$.
Since $\sum_i |T_{ip}-\beta_{ip}|\le \sum_i (T_{ip}+\beta_{ip})=2$,
\[
\Big\|\sum_i (T_{ip}-\beta_{ip})\,R_i\Big\|
\ \le\ 2\cdot \tfrac16\,1.5^3 d^{3/2}\,L_{\mathrm{Hess}}\ \Delta x^3
\ =\ O(\Delta x^3).
\]
Therefore \eqref{eq:weighted-identity} can be bounded as 
\[
\|v_p^{(2\mathrm{hop})}-v_p^{(\beta)}\|
\ \le\ 1.5^2 d\,\|H_v\|_{\infty,\text{patch}_p}\,\Delta x^2 \ +\ O(\Delta x^3),
\]
which proves the stated $O(\Delta x^2)$ bound.
\end{proof}
\end{theorem}

\subsubsection{G2P Velocity Gradient}

\begin{theorem}[G2P $G$ is second order accurate]
\label{thm:g2p-G-twohop-verbose}
Let
\[
G_c:=\sum_i v(x_i)\otimes \nabla w_{ic}\big|_{x_c},\qquad
G_p^{(2\mathrm{hop})}:=\sum_c w_{cp}\,G_c.
\]
Let $C_p^{(\beta)}$ be the $B_2$ least-squares gradient,
\begin{align*}
D_p&:=\sum_i \beta_{ip}(x_p)\,(x_i-x_p)(x_i-x_p)^\top,\\
M_p&:=\sum_i \beta_{ip}(x_p)\,v(x_i)(x_i-x_p)^\top,\\
C_p^{(\beta)}&:=M_p D_p^{-1}.
\end{align*}
Assume $v$ is $C^{2,1}$ on the compact stencil that influences $p$.
Then there exists a dimension-only constant $C_G(d)$ such that
\[
\boxed{\quad
\|\,G_p^{(2\mathrm{hop})}-C_p^{(\beta)}\,\|
\ \le\ C_G(d)\ L_{\mathrm{Hess}}\ \Delta x^2.
\quad}
\]
\begin{proof} At each center $c$,
\[
\sum_i \nabla w_{ic}(x_c)=0,\qquad
\sum_i (x_i-x_c)\otimes \nabla w_{ic}(x_c)=I,
\]
and the support consists of the $2^d$ corner nodes at offsets $\pm \tfrac{h}{2}$ per axis.

\medskip
\emph{Step 1.} Let's show that $G_c$ is a second-order approximation of $\nabla v(x_c)$.
Taylor-expand at the \emph{center} $x_c$:
\begin{align*}
v(x_i)&=v(x_c)+\nabla v(x_c)(x_i-x_c)+\tfrac12 H_v(x_c):(x_i-x_c)^{\otimes 2}+R_{i,c}, \\
\|R_{i,c}\|&\le \tfrac16 L_{\mathrm{Hess}}\|x_i-x_c\|^3.
\end{align*}
Multiply by $\nabla w_{ic}(x_c)$ and sum over $i$:
\begin{align*}
&G_c=\underbrace{\sum_i v(x_c)\otimes \nabla w_{ic}(x_c)}_{=\,0}
+\underbrace{\sum_i \nabla v(x_c)(x_i-x_c)\otimes \nabla w_{ic}(x_c)}_{=\,\nabla v(x_c)} \\
&+\ \tfrac12\sum_i \big(H_v(x_c):(x_i-x_c)^{\otimes2}\big)\otimes \nabla w_{ic}(x_c) +\sum_i R_{i,c}\otimes \nabla w_{ic}(x_c).
\end{align*}
On a tensor-product Q1 stencil, the quadratic tensor vanishes by 1D symmetry on each axis (the two nodes at $\pm \tfrac{h}{2}$ contribute opposite slopes $\pm\tfrac{1}{2h}$),
hence
\begin{align*}
\|G_c-\nabla v(x_c)\|
&\le \sum_i \|R_{i,c}\|\,\|\nabla w_{ic}(x_c)\| \\
\ &\le\ L_{\mathrm{Hess}}\ \Big(\max_i \|x_i-x_c\|^3\Big)\ \Big(\sum_i \|\nabla w_{ic}(x_c)\|\Big).
\end{align*}
Using $\|x_i-x_c\|=\tfrac{\sqrt d}{2} \Delta x$ and $\|\nabla w_{ic}(x_c)\|=\tfrac{\sqrt d}{2 \Delta x}$ componentwise,
there is a dimension-only constant $A_d$ with
\begin{equation}\label{eq:Gc-second-order}
\quad
\|G_c-\nabla v(x_c)\|\ \le\ A_d\ L_{\mathrm{Hess}}\ \Delta x^2.
\quad
\end{equation}

\medskip
\emph{Step 2.} Next we show that averaging $G_c$ to the particle preserves second order.
Average \eqref{eq:Gc-second-order} with $w_{cp}$ (nonnegative, sum to one):
\[
G_p^{(2\mathrm{hop})}
=\sum_c w_{cp}\,\nabla v(x_c)\ +\ O(\Delta x^2).
\]
Now expand $\nabla v$ at $x_p$:
\[
\nabla v(x_c)=\nabla v(x_p)+H_v(x_p)(x_c-x_p)+\tfrac12 \nabla^3 v(\xi_c):(x_c-x_p)^{\otimes2}.
\]
Since $\sum_c w_{cp}=1$ and $\sum_c w_{cp}(x_c-x_p)=0$,
\[
\sum_c w_{cp}\,\nabla v(x_c) = \nabla v(x_p) + O(\Delta x^2).
\]
Therefore
\begin{equation}\label{eq:G2hop-approx-grad}
\quad
\|\,G_p^{(2\mathrm{hop})}-\nabla v(x_p)\,\|\ \le\ B_d\ L_{\mathrm{Hess}}\ \Delta x^2.
\quad
\end{equation}

\medskip
\emph{Step 3.} Similarly, we can see the $B_2$ least-squares gradient is also second order.
Insert the Taylor expansion at $x_p$ into
\(
M_p=\sum_i \beta_{ip}\,v(x_i)(x_i-x_p)^\top
\)
and
\(D_p=\sum_i \beta_{ip}(x_i-x_p)(x_i-x_p)^\top\).
Constants vanish because $\sum_i\beta_{ip}(x_i-x_p)=0$; the linear term gives
\(M_p=\nabla v(x_p)\,D_p + O(\Delta x^4)\)
(the cubic moment of the symmetric $B_2$ stencil cancels).
Since $D_p\simeq \mu_2 \Delta x^2 I$ with a model-dependent $\mu_2>0$,
\begin{equation}\label{eq:beta-ls-second-order}
\quad
\|\,C_p^{(\beta)}-\nabla v(x_p)\,\|\ \le\ B'_d\ L_{\mathrm{Hess}}\ \Delta x^2.
\quad
\end{equation}

\medskip
\emph{Step 4.} Finally, we apply triangle inequality to combine \eqref{eq:G2hop-approx-grad} and \eqref{eq:beta-ls-second-order}:
\begin{align*}
\|G_p^{(2\mathrm{hop})}-C_p^{(\beta)}\|
\ &\le\ \|G_p^{(2\mathrm{hop})}-\nabla v(x_p)\|+\|C_p^{(\beta)}-\nabla v(x_p)\| \\
\ &\le\ (B_d+B'_d)\,L_{\mathrm{Hess}}\,\Delta x^2.
\end{align*}
Setting $C_G(d):=B_d+B'_d$ concludes our proof.
\end{proof}
\end{theorem}

\section{Rotation-Free Stretch Reference}
\label{app:rotfree}

In \autoref{subsec:rotfree} we proposed a rotation-free stretch reference. We claimed that one does not need to have a full deformation gradient on quadrature locations, and the stretch tensor itself is enough for us to define an incremental potential governing grid state evolution. Here we prove that doing so, compared to an imaginary ``exact'' situation where we did have accurate deformation gradients at quadratures, is a second order accurate ($O(\Delta t)^2$) approximation to the  velocity solution.

We work over a single backward Euler step in an updated Lagrangian, velocity-primary formulation on a center grid. At each active quadrature $c$, the previous deformation admits a polar split $F_c^n=R_c\,S_c^{\mathrm{(phys)}}$ with $R_c$ a rotation and $S_c^{\mathrm{(phys)}}$ a positive stretch. Our integrator does not store $R_c$ at centers. Instead, from the resampled center Kirchhoff stress $\tau_c^n$ we reconstruct a stretch-only base $S_c\succ0$ by solving
\begin{equation}
\label{eq:base-def}
P(S_c)\,S_c^{\top}=\tau_c^n,
\end{equation}
and we take $F^{\mathrm{base}}_{\mathrm{drop},c}:=S_c$. For reference, keeping the old rotation would correspond to $F^{\mathrm{base}}_{\mathrm{keep},c}:=R_c\,S_c^{\mathrm{(phys)}}$.
We assume an isotropic hyperelastic density $\psi$ with first Piola stress $P(F)=\partial \psi /\partial F$ and a $C^2$ tangent $\mathrm D P[F]$, so that
\begin{equation}
\label{eq:isotropy}
\psi(Q_1 F Q_2)=W(F),\quad P(Q_1 F Q_2)=Q_1\,P(F)\,Q_2,
\end{equation}
for all $Q_1,Q_2\in SO(d)$. Isotropy ensures that $\psi$ depends on $F$ only through its stretch, and that $P$ and its tangent commute with left/right rotations. In practice, implicit iterates remain in a bounded trust region enforced by standard damping/line-search, so all $C^2$ bounds below are uniform.

The center trial for a nodal velocity field $v$ is
\begin{equation}
\label{eq:trial-F}
F_{\xi,c}(v)=\bigl(I+\Delta t\,G_c(v)\bigr)\,F^{\mathrm{base}}_{\xi,c},
\qquad
G_c(v)=\sum_{i} v_i\otimes \nabla w_{ic},
\end{equation}
where $\xi\in\{\mathrm{keep},\mathrm{drop}\}$ selects $F^{\mathrm{base}}_{\xi,c}$, and $\nabla w_{ic}$ are the constant Q1 shape gradients evaluated at $x_c$.

We minimize the velocity-primary backward Euler potential
\begin{equation}
\label{eq:Phi}
\Phi_{\xi}(v)=\tfrac12\,(v-v^n)^{\top}M(v-v^n)\;+\;\sum_{c}V_{c}^n\,\psi(F_{\xi,c}(v))\;,
\end{equation}
with lumped mass matrix $M=\mathrm{diag}(m_i)$. The first-order optimality reads
\begin{equation}
\label{eq:residual}
g_{\xi}(v)=\nabla\Phi_{\xi}(v)=M(v- v^n)\;+\;\Delta t f^{\mathrm{int}}_{\xi}(v)=0,
\end{equation}
where internal forces are built from the first Piola Kirchoff stress tensor in an updated Lagrangian manner:
\begin{align}
f^{\mathrm{int}}_{\xi,i}(v)&=-\sum_{c}V_{c}^n\,Q_{\xi,c}(v)\,\nabla w_{ic}, \label{eq:fint} \\
Q_{\xi,c}(v):&=P(F_{\xi,c}(v))\,F^{\mathrm{base}\,\top}_{\xi,c}.
\end{align}

Our goal is to compare the two fully assembled steps -- keep-$R$ and drop-$R$ -- and prove that the corresponding solutions $v_{\mathrm{keep}}$ and $v_{\mathrm{drop}}$ satisfy
\begin{equation}
\label{eq:goal}
\|v_{\mathrm{keep}}-v_{\mathrm{drop}}\|=O(\Delta t^2).
\end{equation}
The argument proceeds in four steps: (i) per-quadrature elastic energies agree to first order in $\Delta t$; (ii) per-quadrature implicit stress differ by $O(\Delta t)$ at the same $v$; (iii) hence the residuals $g_{\mathrm{keep}}(v)$ and $g_{\mathrm{drop}}(v)$ differ by $O(\Delta t^2)$ at the same $v$; (iv) residual gap translates into a velocity gap of the same order.

\subsection{Elastic Energies Agree to $O(\Delta t)$}
Fix a quadrature $c$ and suppress $c$ for readability. Consider $\phi_{\xi}(\Delta t;v)=\psi\big(( I+\Delta t\,G(v))\,F^{\mathrm{base}}_{\xi}\big)$. A Taylor expansion at $\Delta t=0$ gives
\begin{align}
\phi_{\xi}(\Delta t;v) 
&= \psi(F^{\mathrm{base}}_{\xi})
+ \Delta t\,P(F^{\mathrm{base}}_{\xi}):\!\big(G(v)\,F^{\mathrm{base}}_{\xi}\big) \nonumber \\
&+ \tfrac12\,\Delta t^2\, Q_{\xi}(v) + O(\Delta t^3). \label{eq:taylor-energy}
\end{align}
By isotropy \eqref{eq:isotropy}, $\psi(RS)=\psi(S)$, so zeroth order matches. For the first variation we use
\[
P(F^{\mathrm{base}}_{\xi}):\!\big(G\,F^{\mathrm{base}}_{\xi}\big)
=\big(P(F^{\mathrm{base}}_{\xi})\,F^{\mathrm{base}\,\top}_{\xi}\big):G
=Q_{\xi}(0):G.
\]
At $v=0$, $F_{\xi}=F^{\mathrm{base}}_{\xi}$ and therefore
\begin{align*}
Q_{\mathrm{keep}}(0)&=P(R\,S^{\mathrm{(phys)}})\,(R\,S^{\mathrm{(phys)}})^{\top}=\tau^n, \\
Q_{\mathrm{drop}}(0)&=P(S)\,S^{\top}=\tau^n,
\end{align*}
by \eqref{eq:base-def}. Hence the linear terms in \eqref{eq:taylor-energy} are identical for keep and drop, and
\begin{equation}
\label{eq:energy-gap}
\bigl|\phi_{\mathrm{keep}}(\Delta t;v)-\phi_{\mathrm{drop}}(\Delta t;v)\bigr|
\le C_E\,\Delta t^2\,\|G(v)\|^2,
\end{equation}
with $C_E$ determined by the $C^2$ bound of $\psi$ on the trust region.

\subsection{Implicit Stress Differ by $O(\Delta t)$}
Expand $P$ at the two bases:
\[
P(F_{\xi})
= P(F^{\mathrm{base}}_{\xi}) \;+\; \mathrm D P[F^{\mathrm{base}}_{\xi}]\!:\!\big(\Delta t\,G\,F^{\mathrm{base}}_{\xi}\big) \;+\; O(\Delta t^2).
\]
Multiplying by $F^{\mathrm{base}\,\top}_{\xi}$ yields
\begin{align}
&Q_{\xi}(v)=P(F_{\xi})\,F^{\mathrm{base}\,\top}_{\xi} \nonumber \\
&=Q_{\xi}(0)\;+\;\Delta t\,\Big(\mathrm D P[F^{\mathrm{base}}_{\xi}]\!:\!(G\,F^{\mathrm{base}}_{\xi})\Big)\,F^{\mathrm{base}\,\top}_{\xi}\;+\;O(\Delta t^2). \label{eq:Q-expansion}
\end{align}
Since $Q_{\mathrm{keep}}(0)=Q_{\mathrm{drop}}(0)=\tau^n$ and $\mathrm D P$ respects \eqref{eq:isotropy}, there exist quadrature-wise constants $C_Q,C'_Q$ such that
\begin{equation}
\label{eq:Q-gap}
\|\,Q_{\mathrm{keep}}(v)-Q_{\mathrm{drop}}(v)\,\|\ \le\ C_Q\,\Delta t\,\|G(v)\|\;+\;C'_Q\,\Delta t^2.
\end{equation}

\subsection{Residuals at the Same $v$ Differ by $O(\Delta t^2)$}
With \eqref{eq:fint} and bounded shape gradients,
\begin{align*}
&\|f^{\mathrm{int}}_{\mathrm{keep}}(v)-f^{\mathrm{int}}_{\mathrm{drop}}(v)\| \\
&\le \sum_{c}V_{c}^n\,\|Q_{\mathrm{keep},c}(v)-Q_{\mathrm{drop},c}(v)\|\,\max_{i}\|\nabla w_{ic}\| \le C_F\,\Delta t,
\end{align*}
so subtracting \eqref{eq:residual} for keep and drop gives the key quantitative statement
\begin{equation}
\label{eq:residual-gap}
\|\,g_{\mathrm{keep}}(v)-g_{\mathrm{drop}}(v)\,\| \;\le\; C_R\,\Delta t^2
\end{equation}
for every $v$ in the trust region.

\subsection{Velocity Difference is $O(\Delta t^{2})$}

We now convert the $O(\Delta t^{2})$ residual gap into an $O(\Delta t^{2})$ velocity gap. Let $g_{\mathrm{keep}}$ and $g_{\mathrm{drop}}$ denote the full step residuals for the keep–$R$ and drop–$R$ variants, respectively, and let $v_{\mathrm{drop}}$ be the drop–$R$ solution:
\[
g_{\mathrm{drop}}(v_{\mathrm{drop}})=0.
\]
We have already showed a uniform residual gap at the same $v$:
\[
\|\,g_{\mathrm{keep}}(v)-g_{\mathrm{drop}}(v)\,\|\ \le\ C_R\,\Delta t^2
\]
for every $v$ in the trust region, hence, evaluating at $v=v_{\mathrm{drop}}$,
\begin{equation}\label{eq:step4-gap-at-drop}
\|\,g_{\mathrm{keep}}(v_{\mathrm{drop}})\,\|\ =\ \|\,g_{\mathrm{keep}}(v_{\mathrm{drop}})-g_{\mathrm{drop}}(v_{\mathrm{drop}})\,\|\ \le\ C_R\,\Delta t^2.
\end{equation}
Let
\[
H_{\mathrm{keep}}(v)\ :=\ \nabla g_{\mathrm{keep}}(v)\ =\ \nabla^2\Phi_{\mathrm{keep}}(v)
\ =\ M\ +\ \nabla^2 E_{\mathrm{keep}}(v),
\]
where $\Phi_{\mathrm{keep}}(v)=\tfrac12\,(v-\hat v)^{\!\top}M\,(v-\hat v)+E_{\mathrm{keep}}(v)$ is the incremental potential, $M\succ0$ is the lumped mass matrix on free DOFs, and $E_{\mathrm{keep}}(v)=\sum_c V_{c}^n\,\psi(F_{\mathrm{keep},c}(v))$ is the elastic energy. Because $F$ depends linearly on $v$ with a prefactor $\Delta t$, two derivatives of $E_{\mathrm{keep}}$ with respect to $v$ bring down $\Delta t^2$:
there exists a constant $C_H$ (uniform on the trust region) such that
\begin{equation}\label{eq:elastic-Hess-bound}
\big\|\,\nabla^2 E_{\mathrm{keep}}(v)\,\big\|\ \le\ C_H\,\Delta t^2\qquad\text{for all $v$ considered.}
\end{equation}
Hence the smallest eigenvalue of $H_{\mathrm{keep}}(v)$ satisfies
\[
\lambda_{\min}\!\big(H_{\mathrm{keep}}(v)\big)\ \ge\ \lambda_{\min}(M)\ -\ \|\nabla^2E_{\mathrm{keep}}(v)\|
\ \ge\ \lambda_{\min}(M)-C_H\,\Delta t^2.
\]
For $\Delta t$ within the step's trust region, we ensure $C_H\,\Delta t^2 \le \tfrac12\,\lambda_{\min}(M)$, and therefore
\begin{equation}\label{eq:alpha}
\lambda_{\min}\!\big(H_{\mathrm{keep}}(v)\big)\ \ge\ \alpha\ :=\ \tfrac12\,\lambda_{\min}(M)\ >\ 0.
\end{equation}
Let $v_{\mathrm{keep}}$ solve $g_{\mathrm{keep}}(v_{\mathrm{keep}})=0$, and set $e:=v_{\mathrm{keep}}-v_{\mathrm{drop}}$.
By the mean-value integral form of Taylor’s theorem,
\begin{equation}\label{eq:mvf}
g_{\mathrm{keep}}(v_{\mathrm{keep}})-g_{\mathrm{keep}}(v_{\mathrm{drop}})
=\Big(\int_0^1 H_{\mathrm{keep}}(v_{\mathrm{drop}}+s\,e)\,ds\Big)\,e.
\end{equation}
Since $g_{\mathrm{keep}}(v_{\mathrm{keep}})=0$, \eqref{eq:mvf} gives identity
\begin{equation}\label{eq:key-identity}
-\,g_{\mathrm{keep}}(v_{\mathrm{drop}})
=\Big(\int_0^1 H_{\mathrm{keep}}(v_{\mathrm{drop}}+s\,e)\,ds\Big)\,e.
\end{equation}
Denote the averaged Jacobian by $\overline H:=\int_0^1 H_{\mathrm{keep}}(v_{\mathrm{drop}}+s\,e)\,ds$. By \eqref{eq:alpha},
$\lambda_{\min}(\overline H)\ge\alpha$, hence
\[
\|e\|\ \le\ \|\overline H^{-1}\|\,\|g_{\mathrm{keep}}(v_{\mathrm{drop}})\|\ \le\ \frac{1}{\alpha}\, \|g_{\mathrm{keep}}(v_{\mathrm{drop}})\|.
\]
Invoking \eqref{eq:step4-gap-at-drop} yields the velocity estimate
\begin{equation}\label{eq:vel-gap}
\boxed{\ \ \|\,v_{\mathrm{keep}}-v_{\mathrm{drop}}\,\|\ \le\ \frac{C_R}{\alpha}\ \Delta t^2,\ \ }
\end{equation}
concluding our proof.

\section{Cardano Solution for 3D Neo-Hookean}
\label{sec:Cardano}

Set the depressed cubic in the standard form \(y^3+p\,y+q=0\) with
\[
p:=S_2,\qquad q:=S_3-J^2 .
\]
The discriminant is
\[
\Delta \;=\;\Big(\frac{q}{2}\Big)^{\!2}+\Big(\frac{p}{3}\Big)^{\!3}.
\]

\begin{itemize}[leftmargin=*]

\item {Case \(\Delta\ge 0\) (one real root):}
\begin{equation}\label{eq:cardano-one-real}
m \;=\; \sqrt[3]{-\frac{q}{2}+\sqrt{\Delta}}\;+\;\sqrt[3]{-\frac{q}{2}-\sqrt{\Delta}}\,, 
\end{equation}
where the cube roots are the {real, sign–preserving} ones, i.e., \(\sqrt[3]{z}=\operatorname{sign}(z)\,|z|^{1/3}\).
\item {Case \(\Delta<0\) (three real roots):} let
\[
r:=2\sqrt{-\frac{p}{3}},\qquad 
\theta:=\frac{1}{3}\arccos\!\left(\frac{3q}{2p}\sqrt{-\frac{3}{p}}\right),
\]
then the three solutions are
\begin{equation}\label{eq:cardano-three-real}
m_k \;=\; r\,\cos\!\Big(\theta-\frac{2\pi k}{3}\Big),\qquad k=0,1,2.
\end{equation}
\end{itemize}

For split Neo–Hookean in 3D we first recover \(J\) from the spherical part,
\[
\alpha \;=\; \tfrac13\operatorname{tr}\tau \;=\; \tfrac{\kappa}{2}(J^2-1) 
\ \Rightarrow\ 
J=\sqrt{\,1+\tfrac{2}{\kappa}\,\alpha\,}\,,
\]
and set the scaling \(s:=\mu\,J^{-2/3}\). The deviatoric offsets are
\[
\delta_i \;=\; \frac{\tau_i-\alpha}{s},\qquad 
S_2=\delta_1\delta_2+\delta_2\delta_3+\delta_3\delta_1,\qquad
S_3=\delta_1\delta_2\delta_3 .
\]
Solve \eqref{eq:cardano-one-real} or \eqref{eq:cardano-three-real} for \(m\), then
\[
\beta_i \;=\; m+\delta_i \ (>0),\qquad 
\sigma_i \;=\; \sqrt{\beta_i}
\]
Choose the (unique) real root that satisfies the {positivity} constraint \(\beta_i>0\) for all \(i\)
(equivalently \(m>-\min_i\delta_i\)); this is the physically admissible branch used in
the stretch reconstruction of \autoref{sec:stretch-recon}.

In practice for numerical stability, we clamp the \(\arccos(\cdot)\) argument in \([-1,1]\). We also usse a real, sign–preserving cube root to avoid spurious complex round–trips. Note that because \(\sum_i\delta_i=0\), one has \(p=S_2\le 0\); thus \(r\) in \eqref{eq:cardano-three-real} is real.

\bibliographystyle{ACM-Reference-Format}
\bibliography{references}

\end{document}